\documentclass[sigconf, nonacm=true,table, xcdraw]{acmart}

\usepackage{graphicx}
\usepackage[shortlabels]{enumitem}
\usepackage{booktabs}

\PassOptionsToPackage{hyphens}{url}

\def\BibTeX{{\rm B\kern-.05em{\sc i\kern-.025em b}\kern-.08emT\kern-.1667em\lower.7ex\hbox{E}\kern-.125emX}}
    
\newcommand{\vinod}[1]{\textcolor{blue}{}}

\usepackage[multiple]{footmisc}

\urldef{\footurl}\url{https://www.accessnow.org/the-toronto-declaration-protecting-the-rights -to-equality-and-non-discrimination-in-machine-learning-systems/}

\usepackage[caption=false]{subfig}

\begin{document}


%

\title{Extending the Machine Learning Abstraction Boundary: \\ A Complex Systems Approach to Incorporate Societal Context}

%


\author{Donald Martin, Jr.}
\affiliation{%
  \institution{Google}
}

\author{Vinodkumar Prabhakaran}
\affiliation{%
  \institution{Google}
}
\author{Jill Kuhlberg}
\affiliation{%
 \institution{System Stars}
 }
\author{Andrew Smart}
\affiliation{%
 \institution{Google}
}
\author{William S. Isaac}
\affiliation{%
 \institution{DeepMind}
}
%




%
\renewcommand{\shortauthors}{Martin et al. 2020}

%
\begin{abstract}


Machine learning (ML) fairness research tends to focus primarily on mathematically-based interventions on often opaque algorithms or models and/or their immediate inputs and outputs. 
Such oversimplified mathematical models abstract away the underlying societal context where ML models are conceived, developed, and ultimately deployed. As fairness itself is a socially constructed concept that originates from that societal context along with the model inputs and the models themselves, a lack of an in-depth understanding of societal context can easily undermine the pursuit of ML fairness. 
In this paper, we outline three new tools to improve the comprehension, identification and representation of societal context.  First, we propose a \textit{complex adaptive systems (CAS)} based model and definition of societal context that will help researchers and product developers to expand the abstraction boundary of ML fairness work to include societal context.  Second, we introduce \textit{collaborative causal theory formation (CCTF)} as a key capability for establishing a sociotechnical frame that incorporates diverse mental models and associated causal theories in modeling the problem and solution space for ML-based products. Finally, we identify \textit{community based system dynamics (CBSD)} as a powerful, transparent and rigorous approach for practicing CCTF during all phases of the ML product development process. We conclude with a discussion of how these systems theoretic approaches to understand the societal context within which sociotechnical systems are embedded can improve the development of fair and inclusive ML-based products.

\end{abstract}

%
%

%


%

%
\settopmatter{printfolios=true}

\maketitle

\section{Introduction}
%




The last decade has seen tremendous growth in the field of artificial intelligence (AI), resulting in renowned scholars and world leaders considering it a critical element of an ongoing fourth industrial/technological revolution \cite{schwab2017fourth,floridi2008artificial}.  In large part this revolution has been driven by recent advancements, such as deep learning,  in machine learning model design and development.  However, as the pace of adoption for these technologies accelerates, so too have concerns regarding the fairness, accountability and ethics of machine learning (ML) models and algorithms both within the academic community \cite{lum2016predict,chouldechova2017fair,green2018fair,buolamwini2018gender,raso2018artificial,hoffmann2019fairness} and among the general public \cite{angwin2016machine}.\footnote{\url{shorturl.at/ouJO4}}\footnote{\url{shorturl.at/bgxCK}}
A growing body of research on machine learning fairness attempts to build fairer machine learning systems, however it has been pointed out that these attempts primarily focus on the algorithms and models, and their immediate inputs and outputs \cite{selbst2018fairness}.
The limitations of this observational, statistical approach, when considering normative, constitutive, process-oriented, socially-constructed concepts such as fairness, equity, and ethics \cite{lamertz2002social}, has been a recurring topic in recent fair-ML research \cite{frontiers2018,eubanks2018automating, whittaker2018ai,richardson2019dirty, hardt2016equality}.


{
The challenge of reconciling abstracted social and political considerations related to technological development is neither novel or limited to machine learning. Human-Computer Interaction (HCI) and Science and Technology studies (STS) scholars have long highlighted the struggle of technologists to identify and incorporate these factors into their development processes \cite{callon1986sociology,law1987technology,mantovani1996social,lin2012expectation,kang2015my}. More recently, ML fairness scholars have argued the current ML system design processes exhibit a bias toward abstracting ``away the social context in which systems will be deployed'' \cite{selbst2018fairness} in pursuit of manageable technical problems. This approach is fraught with ethical risks, as ignoring social factors could potentially lead to further exacerbating or introducing new harms in the social context in which the systems are deployed. However, the fact that researchers interchangeably refer to the concept of \textit{social context} as the ``sociotechnical puzzle'' \cite{selbst2018fairness}, ``complex social reality'' \cite{campolo2017ai} and ``the broader context'' \cite{youtube} illustrates a lack of clarity on what social context is. This lack of clarity contributes to the tendency to abstract away social context during ML system design.

In order to combat the tendency to abstract away social context, this paper seeks to re-frame social context as a socio-cultural layer --- which we will refer to as \textit{societal context} ---  of the complex environment in which all technical systems and the social actors that create and are affected by them, exist and interact. Specifically, we introduce and leverage the multidisciplinary complex adaptive system (CAS) theory to develop a taxonomy model of societal context. 

Next, we leverage the taxonomy model and fundamentals of product development processes to propose the concept of collaborative causal theory formation (CCTF), which we identify as a needed capability for incorporating societal context into the ML system design process in partnership with other (often excluded) stakeholders. We focus particularly on operationalizing CCTF through the use of system dynamics (SD) \cite{forrester1994,sterman2001system}, which is a transparent and rigorous visual and analytical tool for facilitating recursive engagement among diverse stakeholders. In practice, SD and bottom-up variants such as community based system dynamics (CBSD) are analogous to other efforts in the ML fairness community \cite{young2019toward,balaram2018artificial} seeking to aid developers and researchers who desire working as partners with impacted stakeholders to develop greater perspective on the social and ethical dimensions of their research and products.}

\section{Modeling Societal Context}


Incorporating socially constructed concepts such as fairness and ethics into a machine learning system design process requires a deep understanding of the \textit{societal context} within which the system will operate. However, currently, within the ML system development ecosystem there is no concrete definition of societal context, nor a description of its key features and characteristics.  We argue that these elements are prerequisites for developing effective strategies and identifying useful frameworks that researchers and practitioners can leverage to extend the ML system design abstraction boundary to encompass societal context.

To make progress on defining societal context we must choose a perspective from which to think about what society is and what its key elements and characteristics are.  Models are an effective way for communicating, explaining and reasoning about the features and characteristics of complex concepts \cite{page2018model}. 
While the idea of modelling elements of the society has been actively researched for decades \cite{minsky1988society,searle1995construction, bruni2007reassembling}, based on 
the groundwork laid by sociologist Walter Buckley, who asserted in 1968 that society was a complex adaptive system (CAS) \cite{buckley1968modern}, we leverage CAS theory to identify the salient features and characteristics of societal context.

\subsection{Complex Adaptive Systems (CAS)}

Complex Adaptive System theory has its origins in general systems theory \cite{von1950theory} --- which emerged in the 1950s as a cohesive interdisciplinary approach to study systems in all fields of science --- and is a loosely organized field of study often bundled into the broader field of complexity science. Complex adaptive systems are complex in the sense that they are comprised of components that are directly or indirectly related in a causal network, and the behaviour of the system cannot be predicted based solely on the behaviour of its components, and are adaptive in the sense that they adapt to the changes in their environment by mutating or self-organizing their internal structures \cite{cilliers2002complexity, dodder2000complex,holland1995hidden,axelrod2000harnessing,miller2009complex}.
Examples cited in literature for CAS vary from small biological systems such as the cell, the embryo, the brain, and the immune system, to large social systems such as ant colonies, social networks, organizations, and governments.

A detailed exposition of CAS and its various applications is outside the scope of this paper.\footnote{CAS is a large and deep discipline with many aspects we do not address in this paper. For example, self-organization, chaotic behavior, fat tailed behavior and power law scaling - key elements of CAS - are not necessary to detail for the purpose of this paper, which is the introduction of our framework combining elements of CAS and community based system dynamics}
However, the key characteristics of CAS include:
\begin{itemize}[leftmargin=*]
    \item \textbf{Complex}: large number of active elements that continuously interact through information and/or energy exchange.
    \item \textbf{Distributed Control}: individual elements of the CAS are necessarily unaware or oblivious to the system as a whole. Each element interacts with and reacts to its own local environment and is governed by its own rules \cite{cilliers2002complexity}.
    For example, an ant colony is a CAS comprised of many individual ants (CAS in and of themselves); no individual ant has the master plan for the complex nests the colony builds or has knowledge of each individual ants motives or behaviors.
    \item \textbf{Aggregation}: individual elements of CAS combine to form aggregate elements. Aggregated elements at one level of organization become building blocks for emergent aggregate properties at a higher level leading to hierarchical organization\cite{holland1995hidden,holland2012signals}. For example amino acids combine to form proteins, proteins combine to form organelles and so on until ants and ant colonies are ultimately formed.
    \item \textbf{Adaptive}: elements update their structures in order to adapt to the constantly changing environment that results from element interactions.
    \item \textbf{Non-linearity}: interactions between the elements is often non-linear; small changes can have large effects in the system.
    \item \textbf{Feedback loops}: interactions are characterized by feedback loops between elements that can be positive (reinforcing, amplifying) or negative (inhibiting, restraining). 
    \item \textbf{Time delays}: interactions between elements may often involve time  delays; interventions and their impact on the system may not be observed for months or years. 
    \item \textbf{History}: system elements and the overall system have the ability to store state and history so that the past helps to shape present behaviour\cite{majdandzic2014spontaneous}.
    \item \textbf{Stochastic}: the system behavior may be inherently stochastic since each element can have randomness in their inner structures/processes. 
    \item \textbf{Emergence}: The system elements interact in stochastic ways but patterns emerge from these interactions in ways that are counter-intuitive and hard to predict \cite{sterman2000business}.
\end{itemize}

\subsubsection{Key Element Types of CAS}

Although CAS theorists agree that adaptive \textit{agents} are a primary element type of CAS there is no definitive, agreed upon taxonomy of CAS element types.  For the purposes of this paper we have synthesized the various other CAS element types proposed in literature \cite{holland1995hidden,holland2012signals,axelrod2000harnessing} into two additional key element types to complement agents. Specifically we utilize the term \textit{precepts} to encompass ``internal models'' \cite{holland1995hidden,holland2012signals} and ``strategies'' \cite{axelrod2000harnessing}, and the term \textit{artifacts} to encompass ``signals/tags'' \cite{holland1995hidden,holland2012signals} and ``artifacts'' \cite{axelrod2000harnessing}.
Distinguishing these types helps us develop a richer representation for societal context, that for instance, separates objectives of agents, from mechanisms of precepts, and outcomes manifested as artifacts.  However, it is important to realize that these distinctions are not always rigid, since some instances of these element types can incorporate properties of more than one element type. The following definitions for each element type will highlight such instances.


\textbf{Agents}
are the 
``bounded subsystems capable of interior processing'' \cite{holland2012signals,minsky1988society}. Agents can be inorganic and inanimate (e.g. machine learning system) or organic and living. They can be as simple as a thermostat or bacteria, or as complex as an RNA molecule, a robot, or a human being. Aggregations of agents such as an organization, corporation or family are referred to as meta-agents.\footnote{Agents are equivalent to Actors and Meta-agents are equivalent to Actants in Actor Network Theory \cite{bruni2007reassembling}}  
Some agents, such as human beings, are also CAS themselves.

\textbf{Precepts} are the internal rules and structures that \textit{constrain and drive the behavior of agents} and ultimately the overall system the agents comprise. 
Agents autonomously adapting to their environments by changing these internal structures through processes such as acclimatization, learning and self-organization is what makes CAS adaptive. For example in human immune systems, cells are agents whose internal structure consists of DNA that can change in response to its environment via mutations. 
Precepts are often mechanisms for memory and persistence of state \cite{ladyman2013complex} as is the case with DNA which persists the instructions for generating the cells that comprise the human body.  
In general, 
precepts 
are highly complex, mostly invisible and extremely difficult to measure \cite{eberhardt2019biased,eckert2005invisible}.

\textbf{Artifacts}
are the results or manifestations of agent behavior in a CAS. Agent behavior generates, contributes to and changes the environments in which they exist. In other words, artifacts reflect the underlying precepts of agents that are hard to directly observe or measure.  For example an artifact of thermostat behavior would be the increased temperature of the room it is in. 
An artifact of an RNA molecule would be a protein. Artifacts come in many forms including 
ant hills, honey, odors, 
buildings, roads, laws, other agents such as offspring or organizations, 
cellphones, ML systems,\footnote{An ML system can also be classified as agent \cite{rahwan_machine_2019} in its own right.} and economic systems. Some artifacts such as organizations, offspring or economic systems are also CAS and/or agents/meta-agents themselves. Similar to precepts, artifacts can also serve as memory mechanisms --- e.g. in the form of a hieroglyph, capacitor, book, or hard drive.

\subsection{A Taxonomic Model of Societal Context} \label{taxonomy}


It was Sociologist Walter Buckley who first introduced the idea that Society can be thought of as a complex  adaptive system \cite{buckley1968modern}. Since its introduction in the 1960s, CAS has been used to model social systems of varying size and complexity such as supply-chain networks \cite{choi2001supply,surana2005supply} and individual organizations \cite{dooley1997complex,boisot1999organizations,schneider2006organizations}, to health care systems \cite{plsek2001challenge,begun2003health,rouse2008health,benham2010social} and economic systems \cite{anderson2018economy}.
Our choice of CAS as the framework to model societal context in ML Fairness research stems from this rich lineage of its successful applications to model social systems.

In applying the CAS model to societal context, we consider societal context to be primarily constructed by human agents \cite{searle1995construction} engaged in a continuous process of simultaneously satisfying their individual complex needs \cite{maslow1943theory} within their physical and socio-cultural environments and creating, changing and adapting (via their precepts) to that same physical and socio-cultural environment. In other words \textit{human agents simultaneously exist in, interact with and create (via artifacts) societal context}.
The precepts of human agents include lower-level, deeply ingrained structures evolved over eons such as the fusiform gyrus which enables humans to recognize faces \cite{mccarthy1997face} and so called fixed action patterns theorized to drive instinctive behavior \cite{moltz1965contemporary}.
For human agents higher-level precepts, central to socialization and that impact decision making processes, include emotions, learned concepts and patterns such as race and gender role stereotypes, biases, beliefs, attitudes, assumptions, values and self-identity. Higher-level precepts also include the concept of aggregate models of the world --- theorized to be central to reasoning and at the heart of human problem solving, decision making, causal inference and goal-directed counter-factual thinking \cite{epstude2008functional, yeung2011investigation} ---  that various disciplines refer to as mental models \cite{forrester1971counterintuitive, camp2009mental,richardson1994foundations}.



\begin{figure}
  \centering
    \includegraphics[width=0.45\textwidth]{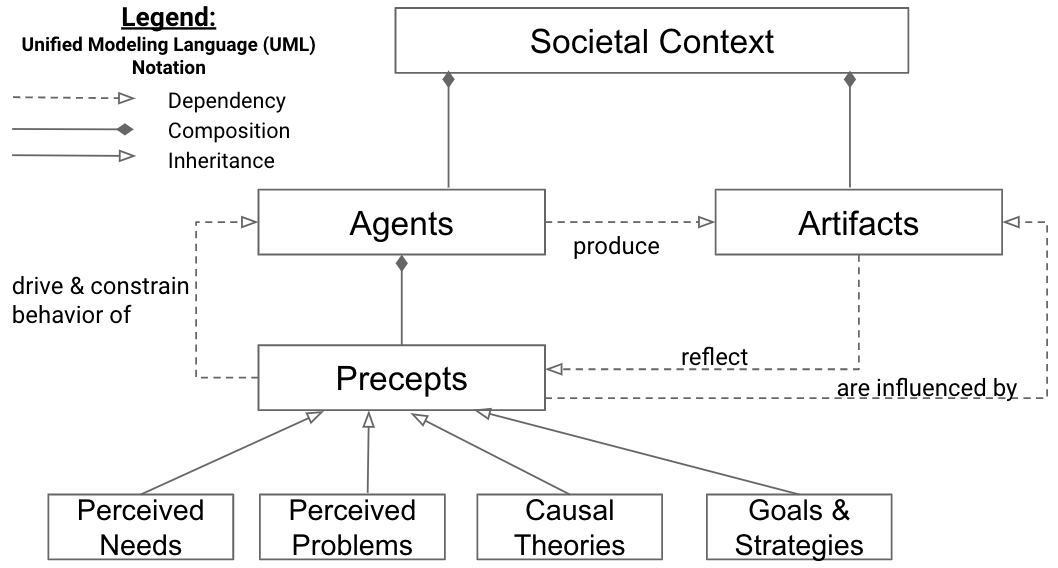}
  \caption{Taxonomic Model of Societal Context using a Complex Adaptive System perspective. \label{sc_model}}
\end{figure}

Figure \ref{sc_model} depicts the taxonomic relationship between the key element types that comprise societal context from a CAS perspective.  Precepts drive and constrain the behavior of agents and are reflected in the artifacts that result from that behavior. In turn, precepts are influenced by the artifacts they are exposed to, resulting in the feedback loops that contribute to the dynamically complex \cite{sterman2000business} nature of societal context. In the following subsections we will highlight key aspects of precepts (\ref{sec_precepts}) and artifacts  (\ref{sec_interventions}) 
that are relevant to ML Fairness and the extension of the ML system design abstraction boundary.



\subsubsection{Four Key Precepts}
\label{sec_precepts}

The primacy of human agents in creating societal context makes their precepts --- which shape what human agents ``see, think and do'' \cite{eberhardt2019biased} --- \textit{the most influential feature of societal context}. As such any approach to extend the abstraction boundary of the ML system design process to include societal context should be centered around identifying and representing human precepts. In particular, the mental models at the foundation of the human decision processes \cite{yeung2011investigation} of the specific humans who fund, build, utilize, and are impacted by the products ML systems are components of should all be considered endogenous to the system being designed. For instance, ignoring the role of human decision processes (e.g. those of judges, defendants and their families) in the case of risk assessment systems will lead to unfair outcomes \cite{selbst2018fairness}. Below we enumerate four key precepts that are essential to uncover the societal context that surrounds ML systems/products.  Although we've enumerated these precepts separately, in reality they are interdependent, overlapping and mutually influential. Additionally, they are supported and influenced by the values, emotions, biases and stereotypes held by the agent.

\begin{enumerate}[leftmargin=*]

\item \textbf{perceived needs}, ranging from fundamental physical needs for food, water and warmth to socially-constructed needs 
such as freedom, safety, fairness, justice and self-actualization \cite{maslow1943theory}.  Satisfying some human agent's (e.g. potential users and the organizations that fund and build products for them) perceived need  is often the motivating factor behind building a product an ML system might be a part of. For example pre-trial risk assessment products (e.g. COMPAS) are intended to address the perceived need to improve pre-trial risk decision making \cite{chiappa2018causal} held by criminal justice organizations.  Perceived needs are very related to and often interchangeable with perceived problems.

\item \textbf{perceived problems}, or the perceived gap between a perceived need and the perceived state of satisfaction of that need. \textit{Perceived problems} can range from an individual human agent's lack of food for their next meal to societal issues
such as homelessness, high rates of crime, immigration or poverty. Societal context (the system of agents, precepts and artifacts) produces patterns of behavior over time (artifacts) such as the growth in the number of homeless people. Some human agents can perceive those artifacts as problems to solve.   Similar to perceived needs, the resolution of a perceived problem is often the motivating factor that drives the development of products,\footnote{Understanding problems as perceived by peripheral stakeholders and social groups is a key analysis factor in the Social Construction of Technology (SCOT) method \cite{pinch1984social}, and is identified as the key method to relate technical artifacts to their ``wider context``.} some of which employ ML systems.

\item \textbf{causal theories}, are the mental models human agents hold about the
structure of cause-to-effect relationships between agents, precepts and artifacts \textit{that cause or lead to a specific problem}.  As we've established earlier, no individual agent`s perception of problem structure can be correct or complete as they are oblivious to and incapable of perceiving and understanding societal context as a whole. In particular, human agents are cognitively incapable of constructing and managing internal causal structures that can take into account the feedback loops and time-delays that characterize problems produced by societal context \cite{forrester1961industrial}. For that reason we refer to these necessarily incomplete perceptions of the causal structure of specific problems as \textit{theories}\cite{tenenbaum2003theory}. Essentially, a human agent`s causal theories are micro-models of societal context. Each human agent's collection of causal-theories is a small patch in the overall fabric of societal context. The causal theories of product managers, ML system designers, potential customers, and potentially impacted peripheral stakeholders are all critical aspects of the societal context that surrounds an ML system.

\item \textbf{goals and strategies} for satisfying needs and/or solving problems \cite{axelrod2000harnessing, holland1995hidden}.  Goals and strategies are closely linked to and often based on causal theories.  As a simple example, 
a job-seeker's causal theory about why they cannot find a job may inform a goal/strategy to move to another state, city or country or enroll in college or a training program. Once this goal/strategy is established, it can become a perceived need to be satisfied or a perceived problem to be solved.
\end{enumerate}


\subsubsection{ML Systems as Artifacts and Interventions} 
\label{sec_interventions}

ML systems, as well as their data inputs and  outputs, are socially constructed artifacts \cite{pinch1984social} of agent behaviors (driven by precepts) that once deployed will become new elements of societal context.  These ML systems and output artifacts can be thought of as \textit{interventions} on some aspect of the system of societal context to solve a perceived problem (precept) or realize a goal (precept) that originated from the human precept aspect of that same societal context.
ML systems are increasingly being used as interventions on societal issues (aka perceived problems) within high-stakes domains such as health and criminal justice. For example, risk assessment and predictive policing systems can be thought of as interventions on the criminal justice system (artifact, meta-agent, and CAS) to solve some perceived problems (precepts) as perceived by a certain set of human agents. As has been demonstrated extensively \cite{ensign2017runaway,lum2016predict}, such interventions can lead to fairness failures when the ML system design abstraction boundary excludes the human precepts and feedback loops that encompass other relevant regions of the societal context.
\section{Collaborative Causal Theory Formation}
\label{sec_extend}


Extending the abstraction boundary of the ML system design process to include societal context is a daunting task. 
The end-to-end Product Development Process (PDP) provides the ``local context'' that shapes abstraction boundary decisions made during ML system design. Hence, a practical first step is to focus on the PDP and the causal theories of the human agents that own, participate and are impacted by it. Here, a ``product'' can be a tool or system developed for internal institutional use or for commercialization and external use by other institutions or individuals.



\subsection{Causal Theories and the Product Development Process}
\label{sec_pdp}
The primary purpose of the PDP is to fulfill the goals and strategies (precepts) of 1) product funders (e.g. finance, sales \& marketing, and product leaders, potential customers) and 2) product owners (e.g. product managers, ML system designers, user experience researchers) by enabling the design and delivery of products that solve perceived problems (precepts) and/or satisfy the perceived needs (precepts) of target stakeholders (agents).  The causal theories (precepts)  of product funders and owners reflect their understanding of the perceived problem they are working to solve and have an enormous impact on how the PDP operates and what it produces. 
As explained in section \ref{sec_precepts}, 
these individual causal theories are necessarily incomplete.

Throughout the PDP, product funders and product owners make high stakes 
design decisions based on their individual, incomplete implicit causal theories about the problem the product is intended to address. These high stakes decisions include deciding what the relevant factors (aka dependent and independent variables) of the problem they have chosen to focus on are and how they are inter-related, who the target stakeholders of the problem solution or product are,  who the peripherally impacted stakeholders are, what  product or sociotechnical system should be deployed to satisfy the need/solve the problem, who should comprise the core product team, whether or not it is appropriate to employ ML to solve the problem and, if ML is chosen, what ML architecture is best suited for the problem at hand. Once the product is deployed it interacts with and influences the perceived needs and problems, goals and strategies and causal theories of target and non-target stakeholders, resulting in feed-back loops.

When incomplete causal theories are the basis for understanding problems and designing product-based solutions, the probability of unintended consequences, sub-optimal solutions and unfair outcomes that negatively impact the most vulnerable stakeholders is likely to increase. For example, incomplete causal theories can lead to excluding peripherally impacted stakeholders (e.g. the family of someone arrested), their perception of problems to be solved, their causal theories about the factors that cause the problem as well as their perspective of what a fair outcome is. In \cite{pinch1984social} the authors point out the important role of the perception of problems by peripheral stakeholders during the development lifecycle of new technologies.  Incomplete causal theories can also lead to excluding the decision processes (driven by precepts) of targeted stakeholders such as the judges who use risk assessment frameworks to inform bail decisions.   Each of these exclusions lead to excluding critical elements of societal context.

More complete causal theories would incorporate the causal theories of target users and peripheral stakeholders and  
will likely lead to a more complete problem understanding, including what interrelated 
factors and feedback loops are most relevant for a given problem, what the most effective interventions could be, what other problems are relevant and what the negative impacts of an intervention could be. 

\subsection{The role of fair-ML researchers and practitioners in the PDP}

As our goal is to identify ways to extend the ML system design abstraction boundary to include societal context, we have chosen to focus on the product development process that envelopes the ML system design and produces ML systems. Fair-ML researchers are typically not owners or drivers of these processes, but rather audit and review the ML system design sub-process outputs --- input datasets, training datasets, model/algorithms, and ML outcomes --- for fairness failures, and develop techniques that ML system designers (a subset of product owners) can utilize for measuring/detecting and mitigating unfair results.  

Due to the limits of purely observational and data analysis approaches to achieving fairness in ML there has been an increasing number of fair-ML researchers delving into the topics of causality \cite{kilbertus2017avoiding,chiappa2019path,chiappa2018causal,madras2019fairness}, feedback loops \cite{ensign2017runaway} and time-delay \cite{liu2018delayed}. A recurring method for representing the causal theories of researchers has been via graphical models, called causal diagrams, of the presumed causal inter-relationships between factors (aka variables) relevant to the problem to be solved or decision to be made. Often times these causal diagrams are constructed using directed acyclic graphs (DAGs).  A number of these approaches leverage Structural Causal Models (SCMs) and Causal Bayesian Networks \cite{madras2019fairness,chiappa2018causal} for measuring unfairness in datasets or building fair decision making models that have biased data as inputs.\footnote{SCMs are optimized for bridging the gap between qualitative and quantitative description and to explicitly deal with the fact that data alone cannot be utilized to understand the underlying causal structures that generated it \cite{pearl2018book}}\footnote{Critiques of counterfactual approaches caution that treating concepts such as race and gender as variables vs complex systems in and of themselves limits their reliability and effectiveness \cite{kohler2018eddie}}
Although these approaches incorporate the concept of causality into ML fairness research, they tend to focus on leveraging those concepts to intervene on ML system inputs or outcomes, not for extending the ML system design abstraction boundary to include societal context.

\subsection{Improving the PDP through Collaborative Causal Theory Formation}
\label{sec_cctf}
As the causal theories of problem funders, product owners, target stakeholders and peripheral stakeholders of the product being designed comprise the core of its societal context, extending the abstraction boundary of the ML System Design process to encompass them requires updating the PDP to overcome four fundamental weaknesses:


\begin{enumerate}[leftmargin=*]
\item   
there is often a lack of diversity on product funding and ownership teams which decreases the richness and variety of the causal theories they produce \cite{campolo2017ai}. 
\textit{Solving for fairness and inclusion requires a deep understanding of unfairness and exclusion}. As such the product funding and ownership teams, whose causal theories and problem understanding drive the PDP, must be as diverse as possible (by race, gender, national origin, socio-economic status, etc.) and include people with deep understanding, through lived-experiences, of unfairness and exclusion.
\item the PDP does not incorporate a systems approach to design that acknowledges and contends with the fact that the products are developed and deployed within a societal context that has the characteristics of a complex adaptive system.
\item the PDP does not make the causal theories of product funding and ownership teams explicit. This prevents them from being tested for validity/completeness and from being improved upon. This often results in decisions being made based on available data, not on what might actually be relevant. 
\item peripheral stakeholders, including policy makers and those belonging to social groups that are most vulnerable to unfair outcomes, are often excluded from meaningfully contributing to the product conception phase of the PDP.
\end{enumerate}

Addressing weakness 1 is beyond the scope of this paper. However, we believe that the value and primacy of diverse causal theories will further validate the ongoing efforts to improve the diversity of product teams in the tech industry. 

In order to address weaknesses 2-4, the typical PDP  must be redesigned to elicit and leverage more comprehensive causal theories in all phases of ML product design, development and deployment, but particularly in the product conception and design phase. Product owners and the fair-ML researchers they partner with must a) adopt a systems-based approach to system design b) become keenly aware of their own \textit{causal theories, including their limitations, about the problem to be intervened on} and c) develop the capability to explicitly surface, share and compare their causal theories with those of other key (obvious and non-obvious) stakeholders. 

In other words, the PDP must incorporate \textbf{the capability to collaboratively discover, understand and synthesize the causal theories of key stakeholders into new, more complete causal theories that more accurately reflect the dynamic complexity 
of the societal context in which the ML-based product (intervention) will ultimately be deployed}.  We will generically call this capability \textit{collaborative causal theory formation (CCTF)}.  Although there are a number of methods \cite{rabiee2004focus, peck1998group, noble1988issue} across disciplines such as anthropology, ethnography, economics and the social sciences, for discovering/eliciting the stories and perspectives of groups, these discipline-specific practices are not typically optimized for collaboratively identifying causal theories or contending with dynamic complexity. The outputs of CCTF are qualitative and quantitative artifacts that contribute to comprehensive problem understanding and causal theory improvement. For optimal effectiveness CCTF should be performed in an open, explicit, multidisciplinary manner
that is optimized to incorporate perspectives from people with lived experiences in the
societal context being investigated. 
To this end, we introduce system dynamics (SD) as a powerful tool to integrate CCTF capabilities in ML system design.

\section{System Dynamics (SD)}
\label{section_sd}

System Dynamics was developed in the 1950s by Jay W. Forrester, who was challenged to see if his experience with control systems and early aircraft simulators could be adapted to produce insights into supply chain dynamics that befuddled those working to manage them \cite{forrester2007system}. The exploratory work was later developed into the computer simulation model detailed in \emph{Industrial Dynamics} \cite{forrester1997industrial}, which illustrated that the puzzling oscillations among inventories and orders were actually the result of the managers' failure to account for the feedback effects of their own decisions \cite{forrester1961industrial}. After applying the modeling approach in the management field, Forrester made the first effort to model social systems in his book Urban Dynamics \citep{ForresterJayW1969Ud}. While this work generated insights about the feedback loops connecting issues related to urban growth and decline, it also led to critiques \citep{baker_2019} that the underlying model reflected and promoted the assumptions, worldviews and causal theories of Forrester and his collaborators while neglecting the same for people with lived experiences in urban settings and alternate political views. 
These critiques resulted in the evolution of participatory group model building techniques that foster incorporation of diverse perspectives \citep{hovmand2014group}. In its now more than 60 years of practice, the SD field has now broadened its application base to
environmental sustainability \cite{hjorth2006navigating,saysel2002environmental}, urban planning \cite{han2009application}, epidemiology \cite{thompsonab2008using}, social welfare \cite{hovmand2009sequence}, education \cite{trani2019strengthening}, and public policy \cite{ghaffarzadegan2011small}.   

\begin{figure*}
  \centering
    \includegraphics[width=.95\textwidth]{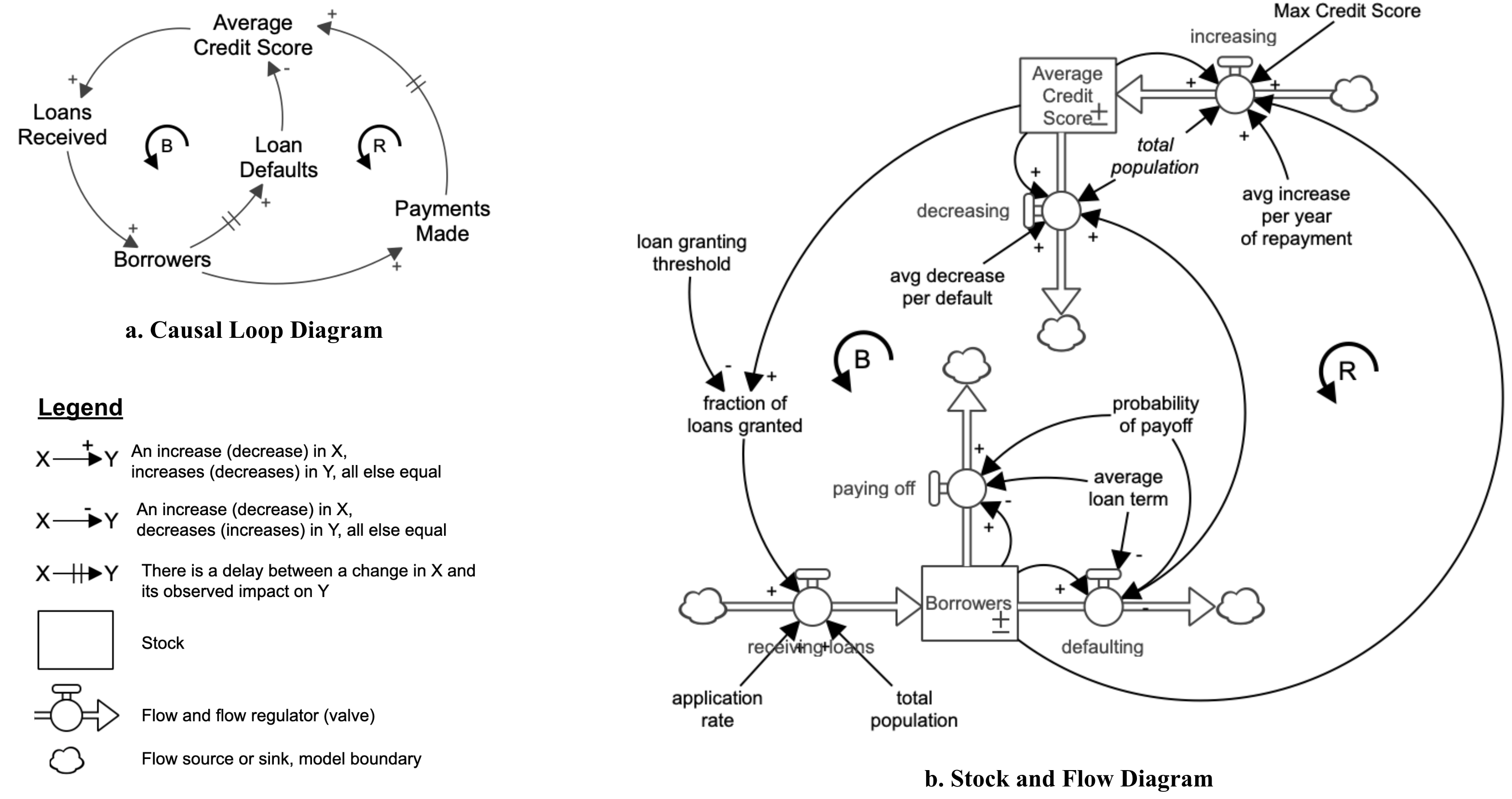}
  \caption{Simple examples of two visual approaches used in system dynamics (causal loop diagrams and stock and flow diagrams) to reflect the hypothesized feedback structures relevant to a simplified Lending System scenario \label{fig_sd_loans_moritz}}
\end{figure*}

SD is defined as \textit{the process of using both informal maps/diagrams and formal models with computer simulation to uncover and understand the dynamics of complex problems from a feedback perspective} \cite{richardson2011reflections}. It is this emphasis on feedback --- reinforcing and balancing processes that unfold over time --- that distinguishes SD from other modeling approaches, and makes it well-suited to incorporate the dynamically complex societal context into the ML system design process. To uncover and understand feedback processes, SD has developed a series of tools that vary in degree of formalism and are designed to provide insight into different aspects of the complex problems they model \cite{richardson2011reflections}. Many of these tools are graphical in nature, requiring modelers to make their causal theories explicit, thereby ensuring transparency \cite{lane2008emergence}.

Recognizing that effective action on a complex problem requires the perspectives, consensus, and coordination of multiple stakeholders, SD has evolved a rich framework to involve stakeholders in the model building process to foster collaboration and learning \cite{kiraly2019dynamics}. 
The reliance on visual tools allow insights and causal theories to be shared and understood with diverse stakeholder groups, enabling a history of participatory approaches \cite{stave2015system,apostolopoulos2018moving,trani2016community, munar2015scaling}.
Community based system dynamics (CBSD) \cite{hovmand2014community} is a particular SD practice approach that engages stakeholders who are embedded in the system of interest to conceptualize the problem, identify the related issues and prioritize interventions based on model supported insights.
These participatory aspects makes a compelling case for SD as a way to perform CCTF in ML Fairness efforts and interventions.


\subsection{Causal Loop Diagrams}

One of the most commonly used visual tools in SD is the causal loop diagram (CLD). The main purpose of the CLD is to show the feedback processes in a system (understood as the set of posited causal structures related to the phenomenon of interest) using a directed graph. Note that CLD is meant to communicate and elicit hypothesized causal relations between variables in a problem space, and is understood to be informal, high-level, and incomplete.

An example of a CLD is shown in Figure~\ref{fig_sd_loans_moritz}a, which offers a simplified representation of a credit score based lending system. The arrows in CLDs represent hypothesized causal links between variables, with the arrowheads and polarity indicating the direction and the nature of influence. Positive polarity represents relationships where an increase (decrease) in one variable triggers an increase (decrease) in the other, all else equal. Negative polarity is used to depict relationships where an increase (decrease) in one variable triggers a decrease (increase) in the other, all else equal.  In the example in Figure~\ref{fig_sd_loans_moritz}a, the relationship between \emph{Payments Made} and \emph{Average Credit Score} is assumed to be of positive polarity since making payments towards debt generally builds credit, \emph{ceteris paribus}, whereas the link between \emph{Loan Defaults} and \emph{Average Credit Score} is negative, since defaulting generally results in score reductions. Any increase (decrease) in the average credit score of a group leads to a corresponding increase (decrease) in the number of loans received by that group, which in-turn  increase (decrease) its borrower pool. 


What distinguishes CLDs from other graphical modeling approaches like DAGs \cite{shrier2008reducing} is that they are designed to capture feedback loops. These loops can be of two types, based on their behavior over time. Reinforcing feedback loops (labeled ``R'' in CLDs) are those processes that amplify system behavior, and can create dynamics of exponential growth or decline. These are often referred to as virtuous or vicious cycles. In turn, balancing feedback loops are those that dampen or counteract change in a system (labeled ``B'' in CLDs). In Figure~\ref{fig_sd_loans_moritz}a, an example of a reinforcing feedback loop is the one generated by the interactions of \textit{Payments Made} and \textit{Average Credit Score} over time: as more borrowers repay their loans, the better their credit scores become, which in turn increases the likelihood of receiving future loans (\textit{Loans Received}). It also illustrates a balancing feedback loop, generated by the interaction of \textit{Loan Defaults} and \textit{Average Credit Score}: as more borrowers default on their loans, the worse their credit becomes, limiting their abilities to qualify for loans (\textit{Loans Received}), and thus their likelihood to default again.  


A further difference between the CLD and other graph based approaches (including those that include feedback like fuzzy cognitive maps \cite{osoba2019beyond}), is its ability to explicitly acknowledge where the relationship between two variables is mediated by the passage of time. These are typically denoted using the same symbol used to represent capacitors in circuit schematics (||).  In Figure~\ref{fig_sd_loans_moritz}a, this means that the impact of repayment on credit score is not only not instantaneous but also that this delay has substantive impacts on the behavior of the system \cite{liu2018delayed}. 

While this deliberately simplified CLD includes only two loops, modelers are encouraged to incorporate as many variables and factors as are required to explain the phenomenon of interest. In particular, CLDs and SD models more generally are not limited to factors for which data is available, and are expected to instead aim to include everything that is relevant (but nothing more) \cite{sterman2000business}. Omitting variables on the basis of lack of quantitative data is explicitly discouraged \cite{forrester1961industrial}; \textit{to assume their effect is zero, is probably the wrongest assumption of them all}. Despite the simplicity of a two loop CLD like the one depicted in Figure~\ref{fig_sd_loans_moritz}a, the complexity of the lending system is still manifest: once the balancing loop is triggered for some populations, the reinforcing loop can turn from a virtuous to a vicious cycle, further limiting their abilities to build credit (as we demonstrate in Section~\ref{sec_sim}).

\subsection{Stock and Flow Diagrams}

A more formal treatment of the causal structures, including the concept of delays and their impact on the system is offered by stock and flow diagrams, perhaps the most commonly used tool in system dynamics.  In addition to representing relationships between variables and feedback loops, stock and flow diagrams require explicit definitions of variables that accumulate, and the precise ways that they accumulate or are depleted over time.  In these diagrams, variables that accumulate are called \textit{stocks} and are drawn as rectangles, and the processes that add to or drain them 
are called flows (inflows and outflows) and are depicted as double-lined/thick arrows or ``pipes'' with valves. The ``clouds'' are the sources and sinks of the flows, and are assumed to have infinite capacity over the time horizon of the model.  These clouds show the model's assumed boundary --- once information or material passes through the flows into a cloud, it ceases to impact the system.\footnote{System dynamics practice encourages ``challenging the clouds''[p. 132] \citep{richmond1993systems}---in other words, critically examining the model's boundary assumptions. Is it appropriate to exclude the stocks currently outside the model boundary? Do those excluded stocks have zero impact on the model? The visualization of stocks and flows supports the discussion of what the the system boundary should include, and simulation modeling provides ways to test the adequacy of assumed boundaries \cite{sterman2000business}.}

Figure~\ref{fig_sd_loans_moritz}b shows a stock and flow representation of the lending system represented in the CLD (Figure~\ref{fig_sd_loans_moritz}a), in which
\emph{Borrowers} and  \emph{Average Credit Score} of the population are now represented as stocks, and are thus assumed to accumulate value over time. The number of borrowers (units = people) accumulates the inflow of people \textit{receiving loans} per year and is depleted by the outflows of people \emph{paying off} the loan completely and \emph{defaulting} on loans per year. In this context the cloud before \emph{receiving loans} indicates the assumption that there is an endless source of individuals who could apply for loans. In turn, in this simplified model, those leaving the system by defaulting or paying off are assumed to not affect the system in any meaningful way,\footnote{the explicit and visual nature of this model boundary decision facilitates input from and discussion by other modelers and stakeholders} and are thus represented as clouds at the ends of the outflows. 
It is important to note that in the process of converting the high-level CLD to a more formal stock and flow diagram, a one-to-one correspondence for the system variables is not enforced; rather the feedback loops are preserved and modeled in more detail. For instance, the causal path from \textit{Borrowers} to the \textit{Average Credit Score} through the  system variable \textit{Payments Made} in the CLD is now represented through a flow (\textit{increasing}) that is guided by the rate at which credit score increases per year as a result of repayments made (\textit{avg increase per year of repayment}), as well as the maximum credit score possible (\textit{Max credit score}). 

The stock and flow diagram is a graphical representation of the system as differential equations that formalizes the behavior of the system over time. The behavior of a stock $S$ can be determined by calculating the integral of its flows over a given time horizon $t$, such that:
\begin{equation} \label{eu_eqn_1}
S(t)=S_0 + \int_{0}^{t} (\text{inflow}(t) - \text{outflow}(t))\;dt
\end{equation}
For instance, in our example, letting  $O$ be the number of borrowers, $r$ be the rate of receiving loans, $f$ be the rate of defaulting (failure to pay), $p$ be the rate of paying the loan off (completely), and $t$ is the unit of time over which the system evolves, the value of $O$ at any given time is defined as  
\begin{equation} \label{eu_eqn_2}
O(t)=O_0 + \int_{0}^{t} (r(t) - p(t)- f(t))\;dt
\end{equation}



Flows and rates are similarly formalized as functions of other elements: stocks, other flows, and exogenous factors. For example, if parameters $x$ and $y$ refer to the \emph{probability of repayment} and \emph{average loan term}, respectively, repaying $p$ is given by
\begin{equation} \label{eu_eqn_3}
p(t)=O(t)\times x/y
\end{equation}

These parameters are determined by the modeler and their explicit definition enhances the transparency of the modeler's causal theory. Together with a set of initial values for the stocks and parameters, this system of equations complete an SD model, and allow modelers to leverage the full power of the stock and flow representation of a system through simulation of its evolution over time. In the absence of empirical estimates, modelers can use values consistent with opinions from relevant stakeholders and conduct sensitivity analyses to understand the ranges of values for which the system presents certain behavior modes --- this is an important aspect of SD that breaks the reliance on only the variables we have data for, and instead provides a way to incorporate qualitative insights into the model building process. 

\subsection{Role of Simulation in SD}
\label{sec_sim}

Despite the usefulness of qualitatively mapping causal theories using CLDs and stocks and flows, the cognitive load required to track the state of the system over time for all but the simplest models is too high \cite{forrester1961industrial, sterman2000business, richardson1991feedback, simon1982models}. Accordingly, numerical simulation approaches have typically been used to study the long-term implications of the causal theory represented in the model. More specifically, simulation is used to test hypotheses about the relationship between the system structure and the behavior it produces, and explore the impacts of parametric and structural changes (e.g., adding/removing feedback loops or flows).


\begin{figure}
  \centering
    \includegraphics[width=.9\linewidth]{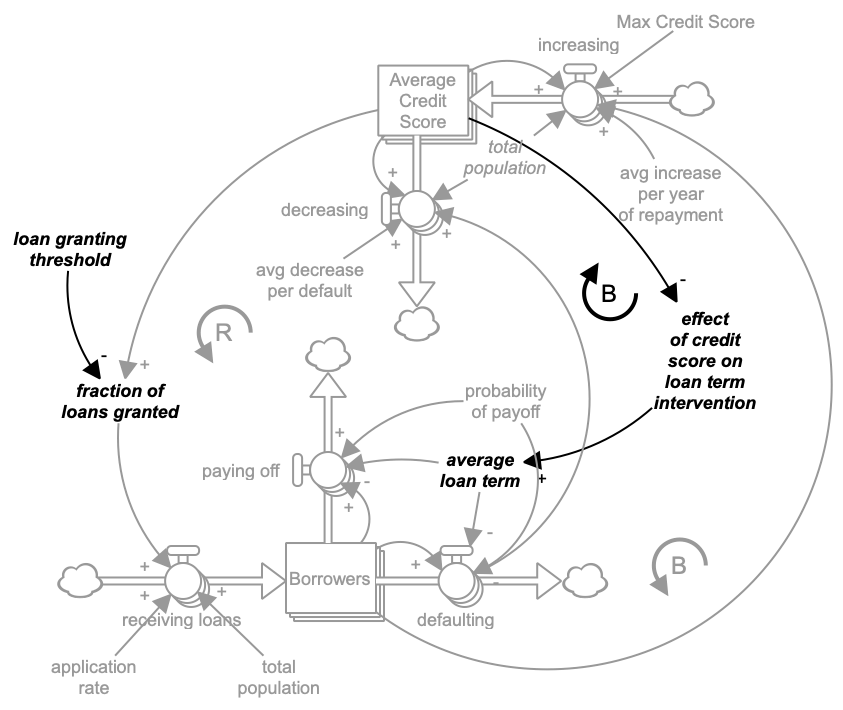}
  \caption{Stock and flow diagram of the lending system highlighting two proposed interventions --- lowering the credit score threshold for granting loans and adjusting the loan term length based on credit score  .\label{fig:SFD_interventions}}
\end{figure}

\begin{figure}
  \centering
    \includegraphics[width=.9\linewidth]{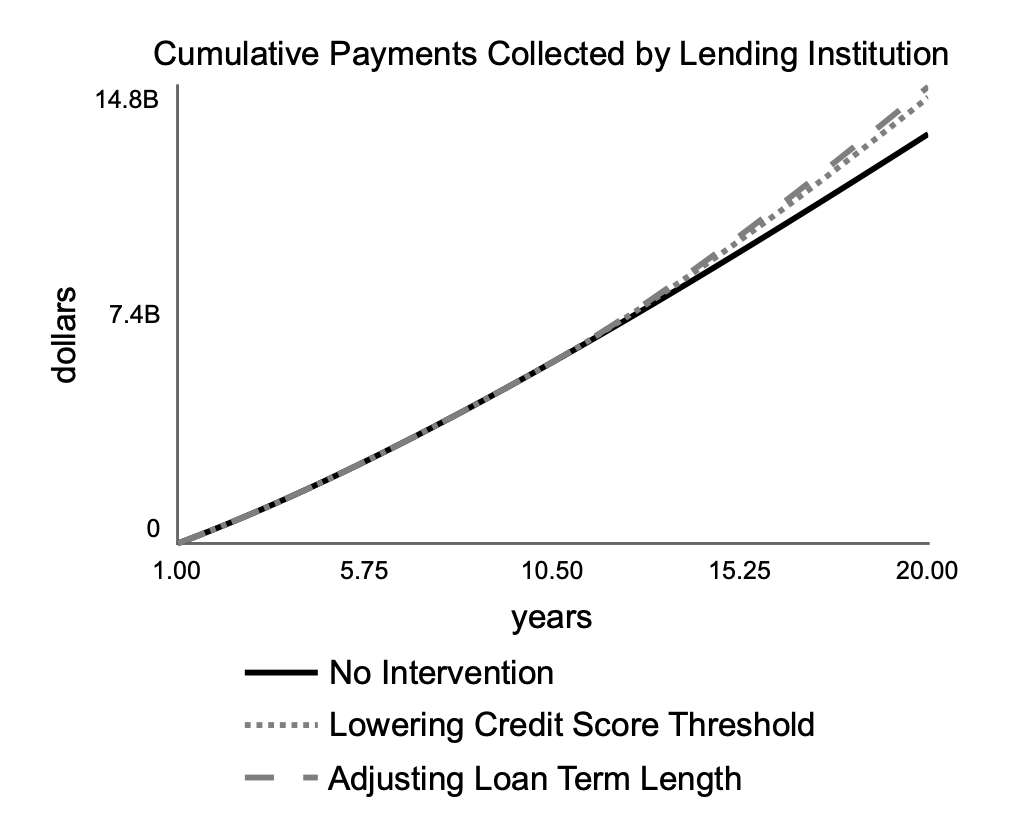}
  \caption{Simulation results comparing a lending institution's cumulative profits under different interventions in the simplified lending system model, assuming an average monthly payment per borrower of \$1000. \label{fig:sim_profits}}
\end{figure}

\begin{figure*}[t]
\centering
\includegraphics[width=.95\linewidth]{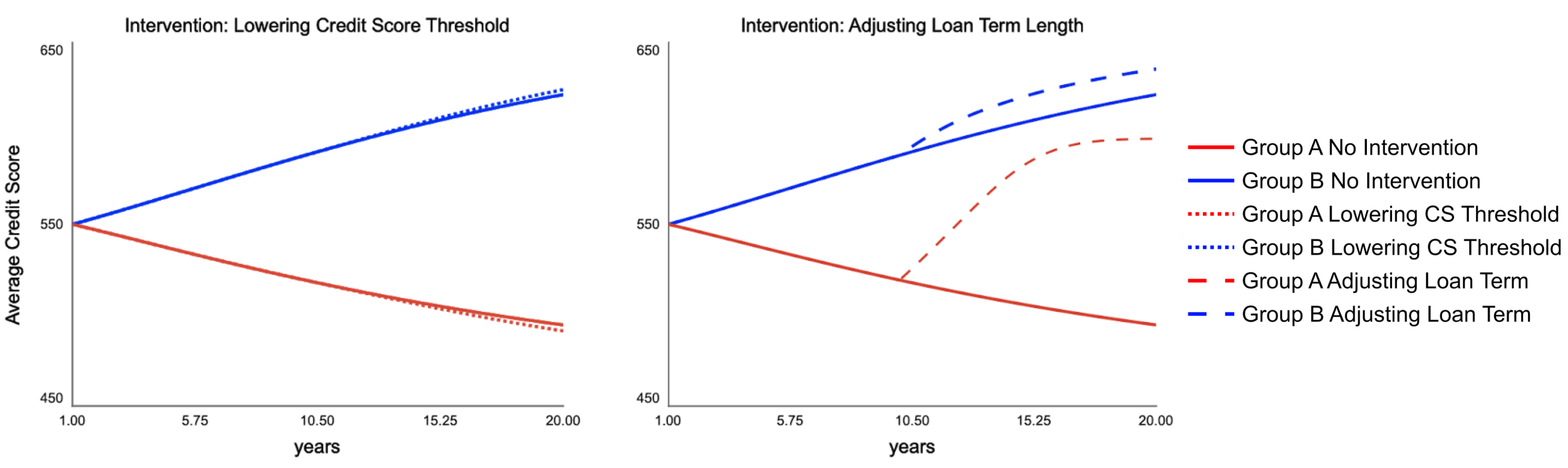}


\caption{Simulation results comparing long-term trajectories of average credit scores for two groups of people (Group A with payoff probability of 0.8 and Group B with payoff probability of 0.6) for different interventions in the lending system. \label{fig:sim_creditscores}}
\end{figure*}


For instance, consider a scenario where a lending institution has a goal (a precept) to improve the profit margin while making more capital available to the markets they serve. 
This could be achieved by designing and implementing an ML-based system that searches the spaces of possible interventions and offers a policy that aligns with this goal. 
The lending institution would employ product funders and owners whose precepts (e.g. goals, strategies, causal theories) drive the PDP that will produce the intervention in question.  
Suppose that such a predictive ML-based system identified two potential strategies that might both improve the profit margin and broaden access to capital. The first lowers the credit score threshold for granting loans to potential borrowers, and the second, provides longer loan terms for borrowers with lower credit scores. These interventions can be integrated into our simplified example stock-flow model, as shown in Figure \ref{fig:SFD_interventions}, where it becomes more apparent that the former introduces a parametric change, and the latter, a structural one (adding a balancing feedback loop to the system). 

SD provides tools to simulate system behavior over time in response to these interventions. For the purpose of this discussion, we set the initial values for all the variables consistent with an average case, and monitor the evolution of the system for 20 years with the interventions implemented at year 10.
Simulation of the two interventions reveals that they both achieve the goal of increasing profits over time for the lending institution. Figure \ref{fig:sim_profits} depicts the lender's profits under a scenario of no intervention (solid-line curve), and the two aforementioned interventions (dotted-line and dashed-line curves). With respect to the lender's profit margin, these interventions produce virtually indistinguishable results. 

However, strategies that can produce seemingly equivalent results with respect to the outcome of immediate interest, may have disparate impacts on different sub-groups of the population --- differences that would be difficult to trace without a more holistic view of the system. 
For instance, from an ML Fairness perspective,
suppose
that we have two sub-groups of people in the population distinguishable only by their payoff probabilities --- $0.8$ for Group A and $0.6$ for Group B.\footnote{For simplicity, the payoff probability encapsulates all of the unobserved factors that can effect a borrowers ability to repay, such as average income, average education levels, etc.} We set the initial values for all other variables for both groups consistent with an average case (for instance, we chose an initial \textit{Average Credit Score} of 550). As before, we monitor the evolution of the system for 20 years with interventions implemented at year 10.

This more holistic view of the system suggests that these seemingly equivalent strategies can have very different impacts on the two sub-groups that form our population. Figure~\ref{fig:sim_creditscores} tracks the evolution of average credit scores for members of those two groups. The left panel compares a scenario of no intervention to lowering the loan granting credit score threshold by 50\% at year 10. The dotted blue line shows an increasing trajectory of the average credit scores for the group with a higher probability of payoff (Group A) under the intervention. However, the same intervention appears to disadvantage the group with the lower probability of payoff (Group B), widening the gap between the two over time. This illustrates the intuition derived from the CLD in Figure~\ref{fig_sd_loans_moritz}b, whereby the reinforcing loop can in fact operate very differently depending on initial conditions, generating undesirable consequences for some groups in the population. The curves representing those same trajectories under the no intervention scenario, depicted in grey, trace the same trajectories so closely that they are barely noticeable. 

Contrast that result to the simulation results of the loan term length intervention, depicted on the right panel of Figure~\ref{fig:sim_creditscores}. Under this intervention, the average credit scores for both groups improve over time, and particularly so for the group that initially had a lower probability of pay off. While the gap remains, the system reequilibrates to a more equitable position than that induced by the alternative strategy.  The credit score trajectories under this new intervention also noticeably depart from the no intervention scenario (depicted in grey) --- illustrating a well-known property of systems, whereby structural changes tend to have more powerful long-term effects than those that simply adjust parameters \citep{meadows1999leverage}.

We can use Figure~\ref{fig:SFD_interventions} to guide the explanation for the trajectories we observe.  The intervention of lowering the credit score threshold for loan granting increases the number of borrowers quickly, however, with each group's probabilities of payoff remaining constant, the number of defaults also increases. This disproportionately impacts the average credit scores of the group with a lower probability of payoff since each default invokes a large credit score penalty.  The intervention adjusting loan term length, however, slows the rate of defaulting, and at the same time allows borrowers more time to increase their credit score as they make monthly payments. While this balancing loop does not change the exogenous probabilities of payoff for each group, it does not allow the system's reinforcing loop to privilege one group and harm another in the presence of the groups' distinct probabilities of payoff. 


Even though this is a simplified representation of the lending system (meant to illustrate the concepts we have introduced throughout the paper), its lessons, with respect to the benefits of incorporating a holistic view of a system, hold in general.  While a predictive ML model can reveal the relationships (as shown in Figure~\ref{fig:sim_creditscores}) between lending thresholds, loan terms and overall profitability of approved loans, only a dynamic feedback perspective can uncover the long-term consequences of activating those relationships, and the mechanisms through which they are likely to operate. As a result it is paramount that these models have a boundary that is extensive enough to encompass the concerns and experiences of those embedded in the system.  


More specifically, when considering proposed interventions, stakeholder groups impacted by the system may identify other (unanticipated) consequences, some of which may imply factors, feedback loops and problematic system behaviors not considered during the causal structure modeling process. 
Observing how feedback loops can change the behavior of a system drastically (as it did in our simple lending system model) highlights the need to partner with diverse stakeholders with unique and relevant causal theories in an iterative modeling process that often begins with establishing a ``reference mode.''  Establishing a ``reference mode'' makes the historical and desired future system behavior explicit and requires the participation of those who stand to be affected by its evolution.




\subsection{Problem Definition with Reference Modes}
In general, graphs like the ones in depicted in Figures~\ref{fig:sim_profits} and \ref{fig:sim_creditscores} are indicative of what those who are modeling the system (typically product funders and owners) believe are important factors to focus on and what they consider to be desirable trajectories over time. Such preferences are not universal and are often not reflective of the precepts (e.g. needs, goals, strategies) of non-target stakeholders. Should all credit scores increase over time? Are increases desirable even if the number of borrowers declines? Should we be focusing on credit scores or financial well-being? What about the profitability of lending institutions? In SD, the process of identifying what the problematic and/or targeted behaviors are is known as defining the \emph{reference mode}. The reference mode is comprised of three components: a clearly defined target/goal or desired-state, a clear idea of the current/problem state, and the clearly depicted gap between the two \cite{saeed1992slicing, repenning2017most}.  It is often expressed as a graph over time, depicting the desired and feared trends of an outcome variable of interest.  These graphs are highly influenced by the goals, strategies and implicit causal theories of their creators.  

In addition to enabling easy comparison to figures resulting from simulation like the ones in Figure~\ref{fig:sim_creditscores}, mapping the reference mode visually promotes dialogue, invites critique and revisions about what is considered the focal variable(s), what the goal pattern of behavior is, and what its associated time horizon is. Reference modes also set the boundaries of the model, as \textit{the goal of SD is to model problems, not the whole system} \cite{sterman2001system}.

While reference modes can be informed by historical data \cite{saeed1992slicing}, care must be taken to avoid choosing focal variables based solely on data availability. To limit reference modes to indicators that already exists runs the risk of conflating what is familiar, tangible, and potentially biased, with what is actually important \cite{sterman2006learning}. 
In a way, reference modes anchor the entire model building process. Not only do they define the dynamics of interest, but also---and perhaps more fundamentally---which variables are most important in the system and which problematic dynamics (e.g. disparate outcome trends) ought to be addressed. Since these decisions determine in large part the kind of insights that can be derived from the modeling exercise, it is of the utmost importance that those likely to be impacted by decisions based on those insights, be partners in the model building process---particularly, in the definition of the reference mode.  In ML fairness, this would imply collaborating with stakeholders to ensure their perspectives and causal theories inform the definition of the reference mode, and in so doing, identifying the most important problems and desired outcomes. In SD practice, this is typically achieved by engaging in participatory modeling, and specifically community based system dynamics.

\subsection{Community Based System Dynamics}

SD has a rich history of involving stakeholders in the model building process to foster collaboration and learning \cite{kiraly2019dynamics}. Community based system dynamics (CBSD) \cite{hovmand2014community} is a particular SD practice approach that engages stakeholders who are embedded in the system of interest to conceptualize the problem, identify the related issues and prioritize interventions based on model supported insights. More than just involving participants in the modeling process to elicit information, CBSD has the explicit goal of building capabilities within communities to use SD and systems thinking tools, distinguishing it from other participatory approaches in SD that often convene participants to gather information from them about an outsider-defined problem to inform an SD model and/or facilitate activities for participants to interact with a simulation model \citep{kiraly2019dynamics}. Building capabilities enables stakeholders to more accurately represent their causal theories in the models, which is especially critical when the stakeholders are from marginalized communities that are not represented in the modeling. In this view, individual and community perspectives on the structures that underlie everyday experiences are valued as valid and necessary sources of data, and community perspectives on the analysis and interpretation of models are essential for realizing the value of the approach. 

Best practices for engaging stakeholders in the process of establishing the reference mode, hypothesizing the causal structure of problems using CLDs and stock and flow diagrams and refining simulation models are documented \cite{hovmand2014community, hovmand2014group}. These activities can be adapted for diverse contexts and support the development of capabilities for collaborative causal theory formation (CCTF).  Overall, CBSD has been shown to be useful in a broad range of problem domains such as maternal and child health \cite{munar2015scaling}, identifying food system vulnerabilities \cite{stave2015system}, mental health interventions \cite{trani2016community} and alcohol abuse \cite{apostolopoulos2018moving}, to name a few. 

In the domain of ML (un)fairness, the practice of CBSD can help center the voices and lived experiences of those marginalized communities that are typically negatively impacted by ML-based products. If the goal is to design fairer ML-based tools and products that do not harm peripheral stakeholders, it is imperative to not only partner with those stakeholders to model the long-term dynamics created by those products when implemented in complex societal contexts, but to also build the capabilities of stakeholders to define and negotiate together what fairness means in those contexts.  

\section{Discussion and Recommendations}


In this paper we add to the argument that in order to effectively evaluate and ensure fair outcomes for ML-based products, we need to expand the ML system design abstraction boundary to include the broader societal context \cite{selbst2018fairness, liu2018delayed}. 
Specifically, we identify and tackle three major weakness (\ref{sec_cctf}) in the typical PDP that impedes the consideration of societal context at scale: 1) lack of systems-based approach to product development and design, 2) lack of methods to transparently articulate and improve the causal theories of product owners that drive the PDP, and 3) limited involvement of stakeholders and communities most negatively impacted by ML-based products. Towards addressing these gaps,
\begin{enumerate}[leftmargin=*]
\item we propose a CAS-based taxonomic model of the key interacting elements (agents, precepts and artifacts) of societal context that ML System designers and fair-ML researchers can use towards extending the abstraction boundary.
\item we introduce collaborative casual theory formation as an essential capability to incorporate diverse stakeholder perspectives into designing fairer ML systems. 
\item we demonstrate system dynamics as a rigorous and scalable approach to model the dynamic complexity that characterizes the societal context in which systems will be deployed and propose CBSD as a means to prioritize incorporating the causal theories of potentially impacted peripheral stakeholders.
\end{enumerate}

Our proposal to employ a complex systems approach to incorporate and understand societal context leans on the rich history of CAS theory and its successful application in a wide array of domains ranging from supply chain networks, health care systems and economic systems. 
SD modeling tools can reflect precisely the characteristics that make societal context dynamically complex, namely feedbacks, accumulations, time delays, and the bounded rationality of agents. While efforts to incorporate causal theories into ML-based products is not new \cite{kilbertus2017avoiding}, we reflect that the failure to incorporate these other characteristics that contribute to the dynamic complexity of societal context into account can have harmful results. As Sterman writes, ``Side effects are not a feature of reality, but a sign that the boundaries of our mental models are too narrow, our time horizons too short'' \cite{sterman2001system}.

Another advantage of our SD-based approach is that it draws heavily on the visual diagramming conventions which emphasize transparency and facilitate the engagement of diverse stakeholders to add, revise and critique causal theories \cite{lane2008emergence, hovmand2014community}. A long lineage of participatory approaches within SD including CBSD and group model building provide evidence of success in developing and using system dynamics models in diverse contexts serve as resources for groups interested in developing SD capabilities in their communities/contexts \cite{munar2015scaling, stave2015system, trani2016community,apostolopoulos2018moving}. Moreover, a strength SD shares with other causal modeling approaches, including Bayesian networks \cite{pearl2009causality, pearl2018book}, is the correspondence between its visualizations and their underlying mathematical representations, which allows stakeholders to do more than visualize, but continue to develop deep insights about important data to collect, consider, and evaluate impact of products and decisions through simulation as well \cite{lane2008emergence, sterman2001system}.

While the methods we have described here can be used to gain a deeper understanding of the societal context associated with a particular problem, we have not described how these methods could be performed at scale by global product companies who have customers in all parts of the world and operate in complex geopolitical environments.  Methods for scaling CCTF efforts, and for managing and leveraging large quantities of qualitative societal context data to support industrial and global scale use cases are areas that require further research. Another set of open questions concern the representation of socially constructed but impactful conceptions such as race and gender in CBSD models as well as in machine learning \citep{keyes2018misgendering,hanna2019towards}. However, partnerships with communities to describe the causal structures of problems that impact them may serve as fertile ground for answering these questions. 




To begin extending the abstraction boundary, it is important first to recognize that given the potentially expansive and authoritative nature of these systems approaches, careful ethical considerations are needed to ensure that the design and deployment of these methods adhere to legal, ethical and moral guidelines.

In addition, we recommend that product owners (particularly product managers and user experience researchers) and fair-ML researchers strive to augment product conception sprints and research initiation brainstorms with CBSD group model building sessions to clarify the space of perceived problems (reference mode \cite{saeed1998defining}) relevant to the product or research effort being conceptualized. The outputs of these sprints will be the transparent, shared and more complete causal theories (aka dynamic hypothesis) about the causal structures that cause the perceived problem. These sessions can be purely qualitative and documented via CLDs and/or stock and flow diagrams and will reveal other perceived problems and stakeholder groups that should, respectively, be considered and fully participate in the next round of theory formation. These dynamic hypotheses are micro-models of the region of societal context most relevant to the product in question and can serve as a critical component of the product requirements/specification document that typically serves as the primary input to the ML system design sub-process.  Initiating ML system design with a more comprehensive understanding of the relevant problem factors and their impacts essentially makes them endogenous to the system and extends the abstraction boundary. These new inputs can also drive the criteria for choosing ML architectures and acquiring appropriate datasets vs. relying solely on the relational inductive biases of individual ML system designers and starting with available datasets. 

While the participation of all keys stakeholders in the ML-based product development process is critical, transformative work in this space begins with centering on community stakeholders and their perspectives on the problems at hand. SD practices in this space must especially protect the perspectives shared by marginalized groups, as models that very precisely and explicitly reflect their vulnerabilities could be exploited. Preventing such exploitation requires working with community stakeholders as partners, not as mere informants whose perspectives (data) are mined to supplement a model based on the causal theories of product funders or owners. Moreover, investments in a CBSD approach to building CCTF capabilities can, over time, generate capabilities and interest in ML within communities currently underrepresented in ML-based product development. Such capabilities and interests can foster an environment in which communities proactively model the problems and social inequities \citep{abebe2020roles} they care about and become full partners in driving the development of holistic and fair solutions.


Product owners and fair-ML researchers motivated to build capabilities and competence in applying CBSD to the task of CCTF should prepare for a painstaking journey. Learning and integrating a systems approach into the existing PDP will require influencing teams and stakeholders who are comfortable with existing approaches.

\subsubsection*{Acknowledgments}
We would like to thank 
Emily Denton,
Ben Hutchinson,
Sean Legassick,
Silvia Chiappa,
Matt Botvinick,
Reena Jana, 
Dierdre Mulligan, and
Deborah Raji
for their valuable feedback on this paper.

\newpage

\bibliographystyle{ACM-Reference-Format}
\bibliography{st_paper}


\begin{thebibliography}{111}


\ifx \showCODEN    \undefined \def \showCODEN     #1{\unskip}     \fi
\ifx \showDOI      \undefined \def \showDOI       #1{#1}\fi
\ifx \showISBNx    \undefined \def \showISBNx     #1{\unskip}     \fi
\ifx \showISBNxiii \undefined \def \showISBNxiii  #1{\unskip}     \fi
\ifx \showISSN     \undefined \def \showISSN      #1{\unskip}     \fi
\ifx \showLCCN     \undefined \def \showLCCN      #1{\unskip}     \fi
\ifx \shownote     \undefined \def \shownote      #1{#1}          \fi
\ifx \showarticletitle \undefined \def \showarticletitle #1{#1}   \fi
\ifx \showURL      \undefined \def \showURL       {\relax}        \fi
\providecommand\bibfield[2]{#2}
\providecommand\bibinfo[2]{#2}
\providecommand\natexlab[1]{#1}
\providecommand\showeprint[2][]{arXiv:#2}

\bibitem[\protect\citeauthoryear{Abebe, Barocas, Kleinberg, Levy, Raghavan, and
  Robinson}{Abebe et~al\mbox{.}}{2020}]%
        {abebe2020roles}
\bibfield{author}{\bibinfo{person}{Rediet Abebe}, \bibinfo{person}{Solon
  Barocas}, \bibinfo{person}{Jon Kleinberg}, \bibinfo{person}{Karen Levy},
  \bibinfo{person}{Manish Raghavan}, {and} \bibinfo{person}{David~G Robinson}.}
  \bibinfo{year}{2020}\natexlab{}.
\newblock \showarticletitle{Roles for computing in social change}. In
  \bibinfo{booktitle}{\emph{Proceedings of the ACM Conference on Fairness,
  Accountability, and Transparency (FAT*)}}. \bibinfo{publisher}{ACM},
  \bibinfo{pages}{252--260}.
\newblock


\bibitem[\protect\citeauthoryear{Anderson}{Anderson}{2018}]%
        {anderson2018economy}
\bibfield{author}{\bibinfo{person}{Philip~W Anderson}.}
  \bibinfo{year}{2018}\natexlab{}.
\newblock \bibinfo{booktitle}{\emph{The economy as an evolving complex
  system}}.
\newblock \bibinfo{publisher}{CRC Press}.
\newblock


\bibitem[\protect\citeauthoryear{Angwin, Larson, Mattu, and Kirchner}{Angwin
  et~al\mbox{.}}{2016}]%
        {angwin2016machine}
\bibfield{author}{\bibinfo{person}{Julia Angwin}, \bibinfo{person}{Jeff
  Larson}, \bibinfo{person}{Surya Mattu}, {and} \bibinfo{person}{Lauren
  Kirchner}.} \bibinfo{year}{2016}\natexlab{}.
\newblock \bibinfo{title}{Machine Bias: there's software used across the
  country to predict future criminals. And it's biased against blacks.
  ProPublica 2016}.
\newblock
\newblock


\bibitem[\protect\citeauthoryear{Apostolopoulos, Lemke, Barry, and
  Lich}{Apostolopoulos et~al\mbox{.}}{2018}]%
        {apostolopoulos2018moving}
\bibfield{author}{\bibinfo{person}{Yorghos Apostolopoulos},
  \bibinfo{person}{Michael~K Lemke}, \bibinfo{person}{Adam~E Barry}, {and}
  \bibinfo{person}{Kristen~Hassmiller Lich}.} \bibinfo{year}{2018}\natexlab{}.
\newblock \showarticletitle{Moving alcohol prevention research forward-Part II:
  new directions grounded in community-based system dynamics modeling}.
\newblock \bibinfo{journal}{\emph{Addiction}} \bibinfo{volume}{113},
  \bibinfo{number}{2} (\bibinfo{year}{2018}), \bibinfo{pages}{363--371}.
\newblock


\bibitem[\protect\citeauthoryear{Axelrod and Cohen}{Axelrod and Cohen}{2000}]%
        {axelrod2000harnessing}
\bibfield{author}{\bibinfo{person}{Robert Axelrod} {and}
  \bibinfo{person}{Michael~D Cohen}.} \bibinfo{year}{2000}\natexlab{}.
\newblock \bibinfo{booktitle}{\emph{Harnessing complexity}}.
\newblock \bibinfo{publisher}{Basic Books}.
\newblock


\bibitem[\protect\citeauthoryear{Baker}{Baker}{2019}]%
        {baker_2019}
\bibfield{author}{\bibinfo{person}{Kevin~T Baker}.}
  \bibinfo{year}{2019}\natexlab{}.
\newblock \bibinfo{title}{Model Metropolis}.
\newblock
\newblock
\urldef\tempurl%
\url{https://logicmag.io/play/model-metropolis/}
\showURL{%
\tempurl}


\bibitem[\protect\citeauthoryear{Balaram, Greenham, and Leonard}{Balaram
  et~al\mbox{.}}{2018}]%
        {balaram2018artificial}
\bibfield{author}{\bibinfo{person}{Brhmie Balaram}, \bibinfo{person}{Tony
  Greenham}, {and} \bibinfo{person}{Jasmine Leonard}.}
  \bibinfo{year}{2018}\natexlab{}.
\newblock \bibinfo{booktitle}{\emph{Artificial Intelligence: Real Public
  Engagement}}.
\newblock \bibinfo{type}{{T}echnical {R}eport}.
\newblock


\bibitem[\protect\citeauthoryear{Begun, Zimmerman, and Dooley}{Begun
  et~al\mbox{.}}{2003}]%
        {begun2003health}
\bibfield{author}{\bibinfo{person}{James~W Begun}, \bibinfo{person}{Brenda
  Zimmerman}, {and} \bibinfo{person}{Kevin Dooley}.}
  \bibinfo{year}{2003}\natexlab{}.
\newblock \showarticletitle{Health care organizations as complex adaptive
  systems}.
\newblock \bibinfo{journal}{\emph{Advances in health care organization theory}}
   \bibinfo{volume}{253} (\bibinfo{year}{2003}), \bibinfo{pages}{288}.
\newblock


\bibitem[\protect\citeauthoryear{Benham-Hutchins and Clancy}{Benham-Hutchins
  and Clancy}{2010}]%
        {benham2010social}
\bibfield{author}{\bibinfo{person}{Marge Benham-Hutchins} {and}
  \bibinfo{person}{Thomas~R Clancy}.} \bibinfo{year}{2010}\natexlab{}.
\newblock \showarticletitle{Social networks as embedded complex adaptive
  systems}.
\newblock \bibinfo{journal}{\emph{JONA: The Journal of Nursing Administration}}
  \bibinfo{volume}{40}, \bibinfo{number}{9} (\bibinfo{year}{2010}),
  \bibinfo{pages}{352--356}.
\newblock


\bibitem[\protect\citeauthoryear{Boisot and Child}{Boisot and Child}{1999}]%
        {boisot1999organizations}
\bibfield{author}{\bibinfo{person}{Max Boisot} {and} \bibinfo{person}{John
  Child}.} \bibinfo{year}{1999}\natexlab{}.
\newblock \showarticletitle{Organizations as adaptive systems in complex
  environments: The case of China}.
\newblock \bibinfo{journal}{\emph{Organization Science}} \bibinfo{volume}{10},
  \bibinfo{number}{3} (\bibinfo{year}{1999}), \bibinfo{pages}{237--252}.
\newblock


\bibitem[\protect\citeauthoryear{Bruni and Teli}{Bruni and Teli}{2007}]%
        {bruni2007reassembling}
\bibfield{author}{\bibinfo{person}{Attila Bruni} {and}
  \bibinfo{person}{Maurizio Teli}.} \bibinfo{year}{2007}\natexlab{}.
\newblock \showarticletitle{Reassembling the social - An introduction to actor
  network theory}.
\newblock \bibinfo{journal}{\emph{Management Learning}} \bibinfo{volume}{38},
  \bibinfo{number}{1} (\bibinfo{year}{2007}), \bibinfo{pages}{121--125}.
\newblock


\bibitem[\protect\citeauthoryear{Buckley}{Buckley}{1968}]%
        {buckley1968modern}
\bibfield{author}{\bibinfo{person}{W.F. Buckley}.}
  \bibinfo{year}{1968}\natexlab{}.
\newblock \bibinfo{booktitle}{\emph{Modern Systems Research for the Behavioral
  Scientist: A Sourcebook}}.
\newblock \bibinfo{publisher}{Aldine}.
\newblock
\showISBNx{9780202300115}
\showLCCN{68019888}
\urldef\tempurl%
\url{https://books.google.com/books?id=KRYFAAAAMAAJ}
\showURL{%
\tempurl}


\bibitem[\protect\citeauthoryear{Buolamwini and Gebru}{Buolamwini and
  Gebru}{2018}]%
        {buolamwini2018gender}
\bibfield{author}{\bibinfo{person}{Joy Buolamwini} {and}
  \bibinfo{person}{Timnit Gebru}.} \bibinfo{year}{2018}\natexlab{}.
\newblock \showarticletitle{Gender shades: Intersectional accuracy disparities
  in commercial gender classification}. In \bibinfo{booktitle}{\emph{Conference
  on fairness, accountability and transparency}}. \bibinfo{pages}{77--91}.
\newblock


\bibitem[\protect\citeauthoryear{Callon}{Callon}{1986}]%
        {callon1986sociology}
\bibfield{author}{\bibinfo{person}{Michel Callon}.}
  \bibinfo{year}{1986}\natexlab{}.
\newblock \showarticletitle{The sociology of an actor-network: The case of the
  electric vehicle}.
\newblock In \bibinfo{booktitle}{\emph{Mapping the dynamics of science and
  technology}}. \bibinfo{publisher}{Springer}, \bibinfo{pages}{19--34}.
\newblock


\bibitem[\protect\citeauthoryear{Camp}{Camp}{2009}]%
        {camp2009mental}
\bibfield{author}{\bibinfo{person}{L~Jean Camp}.}
  \bibinfo{year}{2009}\natexlab{}.
\newblock \showarticletitle{Mental models of privacy and security}.
\newblock \bibinfo{journal}{\emph{IEEE Technology and society magazine}}
  \bibinfo{volume}{28}, \bibinfo{number}{3} (\bibinfo{year}{2009}),
  \bibinfo{pages}{37--46}.
\newblock


\bibitem[\protect\citeauthoryear{Campolo, Sanfilippo, Whittaker, and
  Crawford}{Campolo et~al\mbox{.}}{2017}]%
        {campolo2017ai}
\bibfield{author}{\bibinfo{person}{Alex Campolo}, \bibinfo{person}{Madelyn
  Sanfilippo}, \bibinfo{person}{Meredith Whittaker}, {and}
  \bibinfo{person}{Kate Crawford}.} \bibinfo{year}{2017}\natexlab{}.
\newblock \showarticletitle{AI now 2017 report}.
\newblock \bibinfo{journal}{\emph{AI Now Institute at New York University}}
  (\bibinfo{year}{2017}).
\newblock


\bibitem[\protect\citeauthoryear{Chiappa}{Chiappa}{2019}]%
        {chiappa2019path}
\bibfield{author}{\bibinfo{person}{Silvia Chiappa}.}
  \bibinfo{year}{2019}\natexlab{}.
\newblock \showarticletitle{Path-specific counterfactual fairness}. In
  \bibinfo{booktitle}{\emph{Proceedings of the AAAI Conference on Artificial
  Intelligence}}, Vol.~\bibinfo{volume}{33}. \bibinfo{pages}{7801--7808}.
\newblock


\bibitem[\protect\citeauthoryear{Chiappa and Isaac}{Chiappa and Isaac}{2018}]%
        {chiappa2018causal}
\bibfield{author}{\bibinfo{person}{Silvia Chiappa} {and}
  \bibinfo{person}{William~S Isaac}.} \bibinfo{year}{2018}\natexlab{}.
\newblock \showarticletitle{A Causal Bayesian Networks Viewpoint on Fairness}.
  In \bibinfo{booktitle}{\emph{IFIP International Summer School on Privacy and
  Identity Management}}. Springer, \bibinfo{pages}{3--20}.
\newblock


\bibitem[\protect\citeauthoryear{Choi, Dooley, and Rungtusanatham}{Choi
  et~al\mbox{.}}{2001}]%
        {choi2001supply}
\bibfield{author}{\bibinfo{person}{Thomas~Y Choi}, \bibinfo{person}{Kevin~J
  Dooley}, {and} \bibinfo{person}{Manus Rungtusanatham}.}
  \bibinfo{year}{2001}\natexlab{}.
\newblock \showarticletitle{Supply networks and complex adaptive systems:
  control versus emergence}.
\newblock \bibinfo{journal}{\emph{Journal of operations management}}
  \bibinfo{volume}{19}, \bibinfo{number}{3} (\bibinfo{year}{2001}),
  \bibinfo{pages}{351--366}.
\newblock


\bibitem[\protect\citeauthoryear{Chouldechova}{Chouldechova}{2017}]%
        {chouldechova2017fair}
\bibfield{author}{\bibinfo{person}{Alexandra Chouldechova}.}
  \bibinfo{year}{2017}\natexlab{}.
\newblock \showarticletitle{Fair prediction with disparate impact: A study of
  bias in recidivism prediction instruments}.
\newblock \bibinfo{journal}{\emph{Big data}} \bibinfo{volume}{5},
  \bibinfo{number}{2} (\bibinfo{year}{2017}), \bibinfo{pages}{153--163}.
\newblock


\bibitem[\protect\citeauthoryear{Chouldechova and Roth}{Chouldechova and
  Roth}{2018}]%
        {frontiers2018}
\bibfield{author}{\bibinfo{person}{Alexandra Chouldechova} {and}
  \bibinfo{person}{Aaron Roth}.} \bibinfo{year}{2018}\natexlab{}.
\newblock \showarticletitle{The Frontiers of Fairness in Machine Learning}.
\newblock \bibinfo{journal}{\emph{CoRR}}  \bibinfo{volume}{abs/1810.08810}
  (\bibinfo{year}{2018}).
\newblock
\showeprint[arxiv]{1810.08810}
\urldef\tempurl%
\url{http://arxiv.org/abs/1810.08810}
\showURL{%
\tempurl}


\bibitem[\protect\citeauthoryear{Cilliers}{Cilliers}{2002}]%
        {cilliers2002complexity}
\bibfield{author}{\bibinfo{person}{Paul Cilliers}.}
  \bibinfo{year}{2002}\natexlab{}.
\newblock \bibinfo{booktitle}{\emph{Complexity and postmodernism: Understanding
  complex systems}}.
\newblock \bibinfo{publisher}{routledge}.
\newblock


\bibitem[\protect\citeauthoryear{Dodder and Dare}{Dodder and Dare}{2000}]%
        {dodder2000complex}
\bibfield{author}{\bibinfo{person}{Rebecca Dodder} {and}
  \bibinfo{person}{Robert Dare}.} \bibinfo{year}{2000}\natexlab{}.
\newblock \showarticletitle{Complex adaptive systems and complexity theory:
  inter-related knowledge domains}.
\newblock


\bibitem[\protect\citeauthoryear{Dooley}{Dooley}{1997}]%
        {dooley1997complex}
\bibfield{author}{\bibinfo{person}{Kevin~J Dooley}.}
  \bibinfo{year}{1997}\natexlab{}.
\newblock \showarticletitle{A complex adaptive systems model of organization
  change}.
\newblock \bibinfo{journal}{\emph{Nonlinear dynamics, psychology, and life
  sciences}} \bibinfo{volume}{1}, \bibinfo{number}{1} (\bibinfo{year}{1997}),
  \bibinfo{pages}{69--97}.
\newblock


\bibitem[\protect\citeauthoryear{Eberhardt}{Eberhardt}{2019}]%
        {eberhardt2019biased}
\bibfield{author}{\bibinfo{person}{Jennifer~L. Eberhardt}.}
  \bibinfo{year}{2019}\natexlab{}.
\newblock \bibinfo{booktitle}{\emph{Biased: Uncovering the Hidden Prejudice
  That Shapes What We See, Think, and Do}}.
\newblock \bibinfo{publisher}{Penguin Publishing Group}.
\newblock
\showISBNx{9780735224940}
\showLCCN{2018051011}
\urldef\tempurl%
\url{https://books.google.com/books?id=vpdeDwAAQBAJ}
\showURL{%
\tempurl}


\bibitem[\protect\citeauthoryear{Eckert and Bell}{Eckert and Bell}{2005}]%
        {eckert2005invisible}
\bibfield{author}{\bibinfo{person}{Eileen Eckert} {and}
  \bibinfo{person}{Alexandra Bell}.} \bibinfo{year}{2005}\natexlab{}.
\newblock \showarticletitle{Invisible force: Farmers' mental models and how
  they influence learning and actions}.
\newblock \bibinfo{journal}{\emph{Journal of Extension}} \bibinfo{volume}{43},
  \bibinfo{number}{3} (\bibinfo{year}{2005}), \bibinfo{pages}{1--10}.
\newblock


\bibitem[\protect\citeauthoryear{Ensign, Friedler, Neville, Scheidegger, and
  Venkatasubramanian}{Ensign et~al\mbox{.}}{2017}]%
        {ensign2017runaway}
\bibfield{author}{\bibinfo{person}{Danielle Ensign}, \bibinfo{person}{Sorelle~A
  Friedler}, \bibinfo{person}{Scott Neville}, \bibinfo{person}{Carlos
  Scheidegger}, {and} \bibinfo{person}{Suresh Venkatasubramanian}.}
  \bibinfo{year}{2017}\natexlab{}.
\newblock \showarticletitle{Runaway feedback loops in predictive policing}.
\newblock \bibinfo{journal}{\emph{arXiv preprint arXiv:1706.09847}}
  (\bibinfo{year}{2017}).
\newblock


\bibitem[\protect\citeauthoryear{Epstude and Roese}{Epstude and Roese}{2008}]%
        {epstude2008functional}
\bibfield{author}{\bibinfo{person}{Kai Epstude} {and} \bibinfo{person}{Neal~J
  Roese}.} \bibinfo{year}{2008}\natexlab{}.
\newblock \showarticletitle{The functional theory of counterfactual thinking}.
\newblock \bibinfo{journal}{\emph{Personality and Social Psychology Review}}
  \bibinfo{volume}{12}, \bibinfo{number}{2} (\bibinfo{year}{2008}),
  \bibinfo{pages}{168--192}.
\newblock


\bibitem[\protect\citeauthoryear{Eubanks}{Eubanks}{2018}]%
        {eubanks2018automating}
\bibfield{author}{\bibinfo{person}{Virginia Eubanks}.}
  \bibinfo{year}{2018}\natexlab{}.
\newblock \bibinfo{booktitle}{\emph{Automating inequality: How high-tech tools
  profile, police, and punish the poor}}.
\newblock \bibinfo{publisher}{St. Martin's Press}.
\newblock


\bibitem[\protect\citeauthoryear{Floridi}{Floridi}{2008}]%
        {floridi2008artificial}
\bibfield{author}{\bibinfo{person}{Luciano Floridi}.}
  \bibinfo{year}{2008}\natexlab{}.
\newblock \showarticletitle{Artificial intelligence's new frontier: Artificial
  companions and the fourth revolution}.
\newblock \bibinfo{journal}{\emph{Metaphilosophy}} \bibinfo{volume}{39},
  \bibinfo{number}{4-5} (\bibinfo{year}{2008}), \bibinfo{pages}{651--655}.
\newblock


\bibitem[\protect\citeauthoryear{Forrester}{Forrester}{1961}]%
        {forrester1961industrial}
\bibfield{author}{\bibinfo{person}{Jay~W. Forrester}.}
  \bibinfo{year}{1961}\natexlab{}.
\newblock \showarticletitle{Industrial dynamics. 1961}.
\newblock \bibinfo{journal}{\emph{Pegasus Communications, Waltham, MA}}
  (\bibinfo{year}{1961}).
\newblock


\bibitem[\protect\citeauthoryear{Forrester}{Forrester}{1969}]%
        {ForresterJayW1969Ud}
\bibfield{author}{\bibinfo{person}{Jay~W Forrester}.}
  \bibinfo{year}{1969}\natexlab{}.
\newblock \bibinfo{booktitle}{\emph{Urban dynamics}}.
\newblock \bibinfo{publisher}{M.I.T. Press}, \bibinfo{address}{Cambridge,
  Mass.}
\newblock


\bibitem[\protect\citeauthoryear{Forrester}{Forrester}{1971}]%
        {forrester1971counterintuitive}
\bibfield{author}{\bibinfo{person}{Jay~W. Forrester}.}
  \bibinfo{year}{1971}\natexlab{}.
\newblock \showarticletitle{Counterintuitive behavior of social systems}.
\newblock \bibinfo{journal}{\emph{Technological Forecasting and Social Change}}
   \bibinfo{volume}{3} (\bibinfo{year}{1971}), \bibinfo{pages}{1--22}.
\newblock


\bibitem[\protect\citeauthoryear{Forrester}{Forrester}{1994}]%
        {forrester1994}
\bibfield{author}{\bibinfo{person}{Jay~W. Forrester}.}
  \bibinfo{year}{1994}\natexlab{}.
\newblock \showarticletitle{{System dynamics, systems thinking, and soft OR}}.
\newblock \bibinfo{journal}{\emph{System Dynamics Review}}
  \bibinfo{volume}{10} (\bibinfo{year}{1994}), \bibinfo{pages}{245--256}.
\newblock
\showISSN{1099-1727}
\urldef\tempurl%
\url{https://doi.org/10.1002/sdr.4260100211}
\showDOI{\tempurl}


\bibitem[\protect\citeauthoryear{Forrester}{Forrester}{1997}]%
        {forrester1997industrial}
\bibfield{author}{\bibinfo{person}{Jay~W. Forrester}.}
  \bibinfo{year}{1997}\natexlab{}.
\newblock \showarticletitle{Industrial dynamics}.
\newblock \bibinfo{journal}{\emph{Journal of the Operational Research Society}}
  \bibinfo{volume}{48}, \bibinfo{number}{10} (\bibinfo{year}{1997}),
  \bibinfo{pages}{1037--1041}.
\newblock


\bibitem[\protect\citeauthoryear{Forrester}{Forrester}{2007}]%
        {forrester2007system}
\bibfield{author}{\bibinfo{person}{Jay~W. Forrester}.}
  \bibinfo{year}{2007}\natexlab{}.
\newblock \showarticletitle{System dynamics--a personal view of the first fifty
  years}.
\newblock \bibinfo{journal}{\emph{System Dynamics Review: The Journal of the
  System Dynamics Society}} \bibinfo{volume}{23}, \bibinfo{number}{2-3}
  (\bibinfo{year}{2007}), \bibinfo{pages}{345--358}.
\newblock


\bibitem[\protect\citeauthoryear{Ghaffarzadegan, Lyneis, and
  Richardson}{Ghaffarzadegan et~al\mbox{.}}{2011}]%
        {ghaffarzadegan2011small}
\bibfield{author}{\bibinfo{person}{Navid Ghaffarzadegan}, \bibinfo{person}{John
  Lyneis}, {and} \bibinfo{person}{George~P Richardson}.}
  \bibinfo{year}{2011}\natexlab{}.
\newblock \showarticletitle{How small system dynamics models can help the
  public policy process}.
\newblock \bibinfo{journal}{\emph{System Dynamics Review}}
  \bibinfo{volume}{27}, \bibinfo{number}{1} (\bibinfo{year}{2011}),
  \bibinfo{pages}{22--44}.
\newblock


\bibitem[\protect\citeauthoryear{Green}{Green}{2018}]%
        {green2018fair}
\bibfield{author}{\bibinfo{person}{Ben Green}.}
  \bibinfo{year}{2018}\natexlab{}.
\newblock \showarticletitle{``Fair'' Risk Assessments: A Precarious Approach
  for Criminal Justice Reform}. In \bibinfo{booktitle}{\emph{5th Workshop on
  Fairness, Accountability, and Transparency in Machine Learning}}.
\newblock


\bibitem[\protect\citeauthoryear{Han, Hayashi, Cao, and Imura}{Han
  et~al\mbox{.}}{2009}]%
        {han2009application}
\bibfield{author}{\bibinfo{person}{Ji Han}, \bibinfo{person}{Yoshitsugu
  Hayashi}, \bibinfo{person}{Xin Cao}, {and} \bibinfo{person}{Hidefumi Imura}.}
  \bibinfo{year}{2009}\natexlab{}.
\newblock \showarticletitle{Application of an integrated system dynamics and
  cellular automata model for urban growth assessment: A case study of
  Shanghai, China}.
\newblock \bibinfo{journal}{\emph{Landscape and urban planning}}
  \bibinfo{volume}{91}, \bibinfo{number}{3} (\bibinfo{year}{2009}),
  \bibinfo{pages}{133--141}.
\newblock


\bibitem[\protect\citeauthoryear{Hanna, Denton, Smart, and Smith-Loud}{Hanna
  et~al\mbox{.}}{2019}]%
        {hanna2019towards}
\bibfield{author}{\bibinfo{person}{Alex Hanna}, \bibinfo{person}{Emily Denton},
  \bibinfo{person}{Andrew Smart}, {and} \bibinfo{person}{Jamila Smith-Loud}.}
  \bibinfo{year}{2019}\natexlab{}.
\newblock \showarticletitle{Towards a Critical Race Methodology in Algorithmic
  Fairness}.
\newblock \bibinfo{journal}{\emph{arXiv preprint arXiv:1912.03593}}
  (\bibinfo{year}{2019}).
\newblock


\bibitem[\protect\citeauthoryear{Hardt, Price, Srebro, et~al\mbox{.}}{Hardt
  et~al\mbox{.}}{2016}]%
        {hardt2016equality}
\bibfield{author}{\bibinfo{person}{Moritz Hardt}, \bibinfo{person}{Eric Price},
  \bibinfo{person}{Nati Srebro}, {et~al\mbox{.}}}
  \bibinfo{year}{2016}\natexlab{}.
\newblock \showarticletitle{Equality of opportunity in supervised learning}. In
  \bibinfo{booktitle}{\emph{Advances in neural information processing
  systems}}. \bibinfo{pages}{3315--3323}.
\newblock


\bibitem[\protect\citeauthoryear{Hjorth and Bagheri}{Hjorth and
  Bagheri}{2006}]%
        {hjorth2006navigating}
\bibfield{author}{\bibinfo{person}{Peder Hjorth} {and} \bibinfo{person}{Ali
  Bagheri}.} \bibinfo{year}{2006}\natexlab{}.
\newblock \showarticletitle{Navigating towards sustainable development: A
  system dynamics approach}.
\newblock \bibinfo{journal}{\emph{Futures}} \bibinfo{volume}{38},
  \bibinfo{number}{1} (\bibinfo{year}{2006}), \bibinfo{pages}{74--92}.
\newblock


\bibitem[\protect\citeauthoryear{Hoffmann}{Hoffmann}{2019}]%
        {hoffmann2019fairness}
\bibfield{author}{\bibinfo{person}{Anna~Lauren Hoffmann}.}
  \bibinfo{year}{2019}\natexlab{}.
\newblock \showarticletitle{Where fairness fails: On data, algorithms, and the
  limits of antidiscrimination discourse}.
\newblock \bibinfo{journal}{\emph{Under review with Information, Communication,
  and Society}} (\bibinfo{year}{2019}).
\newblock


\bibitem[\protect\citeauthoryear{Holland}{Holland}{1995}]%
        {holland1995hidden}
\bibfield{author}{\bibinfo{person}{John~Henry Holland}.}
  \bibinfo{year}{1995}\natexlab{}.
\newblock \bibinfo{booktitle}{\emph{Hidden order: how adaptation builds
  complexity}}.
\newblock Number 003.7 H6.
\newblock


\bibitem[\protect\citeauthoryear{Holland}{Holland}{2012}]%
        {holland2012signals}
\bibfield{author}{\bibinfo{person}{John~H Holland}.}
  \bibinfo{year}{2012}\natexlab{}.
\newblock \bibinfo{booktitle}{\emph{Signals and boundaries: Building blocks for
  complex adaptive systems}}.
\newblock \bibinfo{publisher}{Mit Press}.
\newblock


\bibitem[\protect\citeauthoryear{Hovmand}{Hovmand}{2014a}]%
        {hovmand2014community}
\bibfield{author}{\bibinfo{person}{Peter~S Hovmand}.}
  \bibinfo{year}{2014}\natexlab{a}.
\newblock \bibinfo{booktitle}{\emph{Community Based System Dynamics}}.
\newblock \bibinfo{publisher}{Springer}.
\newblock


\bibitem[\protect\citeauthoryear{Hovmand}{Hovmand}{2014b}]%
        {hovmand2014group}
\bibfield{author}{\bibinfo{person}{Peter~S Hovmand}.}
  \bibinfo{year}{2014}\natexlab{b}.
\newblock \showarticletitle{Group model building and community-based system
  dynamics process}.
\newblock In \bibinfo{booktitle}{\emph{Community Based System Dynamics}}.
  \bibinfo{publisher}{Springer}, \bibinfo{pages}{17--30}.
\newblock


\bibitem[\protect\citeauthoryear{Hovmand and Ford}{Hovmand and Ford}{2009}]%
        {hovmand2009sequence}
\bibfield{author}{\bibinfo{person}{Peter~S Hovmand} {and}
  \bibinfo{person}{David~N Ford}.} \bibinfo{year}{2009}\natexlab{}.
\newblock \showarticletitle{Sequence and timing of three community
  interventions to domestic violence}.
\newblock \bibinfo{journal}{\emph{American journal of community psychology}}
  \bibinfo{volume}{44}, \bibinfo{number}{3-4} (\bibinfo{year}{2009}),
  \bibinfo{pages}{261}.
\newblock


\bibitem[\protect\citeauthoryear{Kang, Dabbish, Fruchter, and Kiesler}{Kang
  et~al\mbox{.}}{2015}]%
        {kang2015my}
\bibfield{author}{\bibinfo{person}{Ruogu Kang}, \bibinfo{person}{Laura
  Dabbish}, \bibinfo{person}{Nathaniel Fruchter}, {and} \bibinfo{person}{Sara
  Kiesler}.} \bibinfo{year}{2015}\natexlab{}.
\newblock \showarticletitle{``My Data Just Goes Everywhere:'' User Mental
  Models of the Internet and Implications for Privacy and Security}. In
  \bibinfo{booktitle}{\emph{Eleventh Symposium On Usable Privacy and Security
  ($\{$SOUPS$\}$ 2015)}}. \bibinfo{pages}{39--52}.
\newblock


\bibitem[\protect\citeauthoryear{Keyes}{Keyes}{2018}]%
        {keyes2018misgendering}
\bibfield{author}{\bibinfo{person}{Os Keyes}.} \bibinfo{year}{2018}\natexlab{}.
\newblock \showarticletitle{The misgendering machines: Trans/{HCI} implications
  of automatic gender recognition}.
\newblock \bibinfo{journal}{\emph{Proceedings of the ACM on Human-Computer
  Interaction}} \bibinfo{volume}{2}, \bibinfo{number}{CSCW}
  (\bibinfo{year}{2018}), \bibinfo{pages}{88}.
\newblock


\bibitem[\protect\citeauthoryear{Kilbertus, Carulla, Parascandolo, Hardt,
  Janzing, and Sch{\"o}lkopf}{Kilbertus et~al\mbox{.}}{2017}]%
        {kilbertus2017avoiding}
\bibfield{author}{\bibinfo{person}{Niki Kilbertus},
  \bibinfo{person}{Mateo~Rojas Carulla}, \bibinfo{person}{Giambattista
  Parascandolo}, \bibinfo{person}{Moritz Hardt}, \bibinfo{person}{Dominik
  Janzing}, {and} \bibinfo{person}{Bernhard Sch{\"o}lkopf}.}
  \bibinfo{year}{2017}\natexlab{}.
\newblock \showarticletitle{Avoiding discrimination through causal reasoning}.
  In \bibinfo{booktitle}{\emph{Advances in Neural Information Processing
  Systems}}. \bibinfo{pages}{656--666}.
\newblock


\bibitem[\protect\citeauthoryear{Kir{\'a}ly and Miskolczi}{Kir{\'a}ly and
  Miskolczi}{2019}]%
        {kiraly2019dynamics}
\bibfield{author}{\bibinfo{person}{G{\'a}bor Kir{\'a}ly} {and}
  \bibinfo{person}{P{\'e}ter Miskolczi}.} \bibinfo{year}{2019}\natexlab{}.
\newblock \showarticletitle{Dynamics of participation: System dynamics and
  participation--An empirical review}.
\newblock \bibinfo{journal}{\emph{Systems Research and Behavioral Science}}
  \bibinfo{volume}{36}, \bibinfo{number}{2} (\bibinfo{year}{2019}),
  \bibinfo{pages}{199--210}.
\newblock


\bibitem[\protect\citeauthoryear{Kohler-Hausmann}{Kohler-Hausmann}{2018}]%
        {kohler2018eddie}
\bibfield{author}{\bibinfo{person}{Issa Kohler-Hausmann}.}
  \bibinfo{year}{2018}\natexlab{}.
\newblock \showarticletitle{Eddie Murphy and the Dangers of Counterfactual
  Causal Thinking About Detecting Racial Discrimination}.
\newblock \bibinfo{journal}{\emph{Available at SSRN 3050650}}
  (\bibinfo{year}{2018}).
\newblock


\bibitem[\protect\citeauthoryear{Ladyman, Lambert, and Wiesner}{Ladyman
  et~al\mbox{.}}{2013}]%
        {ladyman2013complex}
\bibfield{author}{\bibinfo{person}{James Ladyman}, \bibinfo{person}{James
  Lambert}, {and} \bibinfo{person}{Karoline Wiesner}.}
  \bibinfo{year}{2013}\natexlab{}.
\newblock \showarticletitle{What is a complex system?}
\newblock \bibinfo{journal}{\emph{European Journal for Philosophy of Science}}
  \bibinfo{volume}{3}, \bibinfo{number}{1} (\bibinfo{year}{2013}),
  \bibinfo{pages}{33--67}.
\newblock


\bibitem[\protect\citeauthoryear{Lamertz}{Lamertz}{2002}]%
        {lamertz2002social}
\bibfield{author}{\bibinfo{person}{Kai Lamertz}.}
  \bibinfo{year}{2002}\natexlab{}.
\newblock \showarticletitle{The social construction of fairness: Social
  influence and sense making in organizations}.
\newblock \bibinfo{journal}{\emph{Journal of Organizational Behavior}}
  \bibinfo{volume}{23}, \bibinfo{number}{1} (\bibinfo{year}{2002}),
  \bibinfo{pages}{19--37}.
\newblock


\bibitem[\protect\citeauthoryear{Lane}{Lane}{2008}]%
        {lane2008emergence}
\bibfield{author}{\bibinfo{person}{David~C Lane}.}
  \bibinfo{year}{2008}\natexlab{}.
\newblock \showarticletitle{The emergence and use of diagramming in system
  dynamics: a critical account}.
\newblock \bibinfo{journal}{\emph{Systems Research and Behavioral Science: The
  Official Journal of the International Federation for Systems Research}}
  \bibinfo{volume}{25}, \bibinfo{number}{1} (\bibinfo{year}{2008}),
  \bibinfo{pages}{3--23}.
\newblock


\bibitem[\protect\citeauthoryear{Law et~al\mbox{.}}{Law et~al\mbox{.}}{1987}]%
        {law1987technology}
\bibfield{author}{\bibinfo{person}{John Law} {et~al\mbox{.}}}
  \bibinfo{year}{1987}\natexlab{}.
\newblock \showarticletitle{Technology and heterogeneous engineering: The case
  of Portuguese expansion}.
\newblock \bibinfo{journal}{\emph{The social construction of technological
  systems: New directions in the sociology and history of technology}}
  \bibinfo{volume}{1} (\bibinfo{year}{1987}), \bibinfo{pages}{1--134}.
\newblock


\bibitem[\protect\citeauthoryear{Lin, Amini, Hong, Sadeh, Lindqvist, and
  Zhang}{Lin et~al\mbox{.}}{2012}]%
        {lin2012expectation}
\bibfield{author}{\bibinfo{person}{Jialiu Lin}, \bibinfo{person}{Shahriyar
  Amini}, \bibinfo{person}{Jason~I Hong}, \bibinfo{person}{Norman Sadeh},
  \bibinfo{person}{Janne Lindqvist}, {and} \bibinfo{person}{Joy Zhang}.}
  \bibinfo{year}{2012}\natexlab{}.
\newblock \showarticletitle{Expectation and purpose: understanding users'
  mental models of mobile app privacy through crowdsourcing}. In
  \bibinfo{booktitle}{\emph{Proceedings of the 2012 ACM conference on
  ubiquitous computing}}. ACM, \bibinfo{pages}{501--510}.
\newblock


\bibitem[\protect\citeauthoryear{Liu, Dean, Rolf, Simchowitz, and Hardt}{Liu
  et~al\mbox{.}}{2018}]%
        {liu2018delayed}
\bibfield{author}{\bibinfo{person}{Lydia~T Liu}, \bibinfo{person}{Sarah Dean},
  \bibinfo{person}{Esther Rolf}, \bibinfo{person}{Max Simchowitz}, {and}
  \bibinfo{person}{Moritz Hardt}.} \bibinfo{year}{2018}\natexlab{}.
\newblock \showarticletitle{Delayed impact of fair machine learning}.
\newblock \bibinfo{journal}{\emph{arXiv preprint arXiv:1803.04383}}
  (\bibinfo{year}{2018}).
\newblock


\bibitem[\protect\citeauthoryear{Lum, Bender, and Wilkerson}{Lum
  et~al\mbox{.}}{[n.d.]}]%
        {youtube}
\bibfield{author}{\bibinfo{person}{Kristian Lum}, \bibinfo{person}{Elizabeth
  Bender}, {and} \bibinfo{person}{Wilkerson}.}
  \bibinfo{year}{[n.d.]}\natexlab{}.
\newblock \bibinfo{booktitle}{\emph{FAT* 2018 Translation Tutorial:
  Understanding the Context and Consequences of Pre-trial Detention}}.
\newblock Association for Computing Machinery.
\newblock
\urldef\tempurl%
\url{https://www.youtube.com/watch?v=hEThGT-_5ho}
\showURL{%
\tempurl}


\bibitem[\protect\citeauthoryear{Lum and Isaac}{Lum and Isaac}{2016}]%
        {lum2016predict}
\bibfield{author}{\bibinfo{person}{Kristian Lum} {and} \bibinfo{person}{William
  Isaac}.} \bibinfo{year}{2016}\natexlab{}.
\newblock \showarticletitle{To predict and serve?}
\newblock \bibinfo{journal}{\emph{Significance}} \bibinfo{volume}{13},
  \bibinfo{number}{5} (\bibinfo{year}{2016}), \bibinfo{pages}{14--19}.
\newblock


\bibitem[\protect\citeauthoryear{Madras, Creager, Pitassi, and Zemel}{Madras
  et~al\mbox{.}}{2019}]%
        {madras2019fairness}
\bibfield{author}{\bibinfo{person}{David Madras}, \bibinfo{person}{Elliot
  Creager}, \bibinfo{person}{Toniann Pitassi}, {and} \bibinfo{person}{Richard
  Zemel}.} \bibinfo{year}{2019}\natexlab{}.
\newblock \showarticletitle{Fairness through Causal Awareness: Learning Causal
  Latent-Variable Models for Biased Data}. In
  \bibinfo{booktitle}{\emph{Proceedings of the Conference on Fairness,
  Accountability, and Transparency}}. ACM, \bibinfo{pages}{349--358}.
\newblock


\bibitem[\protect\citeauthoryear{Majdandzic, Podobnik, Buldyrev, Kenett,
  Havlin, and Stanley}{Majdandzic et~al\mbox{.}}{2014}]%
        {majdandzic2014spontaneous}
\bibfield{author}{\bibinfo{person}{Antonio Majdandzic}, \bibinfo{person}{Boris
  Podobnik}, \bibinfo{person}{Sergey~V Buldyrev}, \bibinfo{person}{Dror~Y
  Kenett}, \bibinfo{person}{Shlomo Havlin}, {and} \bibinfo{person}{H~Eugene
  Stanley}.} \bibinfo{year}{2014}\natexlab{}.
\newblock \showarticletitle{Spontaneous recovery in dynamical networks}.
\newblock \bibinfo{journal}{\emph{Nature Physics}} \bibinfo{volume}{10},
  \bibinfo{number}{1} (\bibinfo{year}{2014}), \bibinfo{pages}{34--38}.
\newblock


\bibitem[\protect\citeauthoryear{Mantovani}{Mantovani}{1996}]%
        {mantovani1996social}
\bibfield{author}{\bibinfo{person}{Giuseppe Mantovani}.}
  \bibinfo{year}{1996}\natexlab{}.
\newblock \showarticletitle{Social context in HCl: A new framework for mental
  models, cooperation, and communication}.
\newblock \bibinfo{journal}{\emph{Cognitive Science}} \bibinfo{volume}{20},
  \bibinfo{number}{2} (\bibinfo{year}{1996}), \bibinfo{pages}{237--269}.
\newblock


\bibitem[\protect\citeauthoryear{Maslow}{Maslow}{1943}]%
        {maslow1943theory}
\bibfield{author}{\bibinfo{person}{Abraham~H Maslow}.}
  \bibinfo{year}{1943}\natexlab{}.
\newblock \showarticletitle{A theory of human motivation.}
\newblock \bibinfo{journal}{\emph{Psychological review}} \bibinfo{volume}{50},
  \bibinfo{number}{4} (\bibinfo{year}{1943}), \bibinfo{pages}{370}.
\newblock


\bibitem[\protect\citeauthoryear{McCarthy, Puce, Gore, and Allison}{McCarthy
  et~al\mbox{.}}{1997}]%
        {mccarthy1997face}
\bibfield{author}{\bibinfo{person}{Gregory McCarthy}, \bibinfo{person}{Aina
  Puce}, \bibinfo{person}{John~C Gore}, {and} \bibinfo{person}{Truett
  Allison}.} \bibinfo{year}{1997}\natexlab{}.
\newblock \showarticletitle{Face-specific processing in the human fusiform
  gyrus}.
\newblock \bibinfo{journal}{\emph{Journal of cognitive neuroscience}}
  \bibinfo{volume}{9}, \bibinfo{number}{5} (\bibinfo{year}{1997}),
  \bibinfo{pages}{605--610}.
\newblock


\bibitem[\protect\citeauthoryear{Meadows}{Meadows}{1999}]%
        {meadows1999leverage}
\bibfield{author}{\bibinfo{person}{Donella Meadows}.}
  \bibinfo{year}{1999}\natexlab{}.
\newblock \bibinfo{booktitle}{\emph{Leverage points: Places to intervene in a
  system}}.
\newblock \bibinfo{publisher}{The Sustainability Institute Hartland, VT}.
\newblock


\bibitem[\protect\citeauthoryear{Miller and Page}{Miller and Page}{2009}]%
        {miller2009complex}
\bibfield{author}{\bibinfo{person}{John~H Miller} {and}
  \bibinfo{person}{Scott~E Page}.} \bibinfo{year}{2009}\natexlab{}.
\newblock \bibinfo{booktitle}{\emph{Complex adaptive systems: An introduction
  to computational models of social life}}. Vol.~\bibinfo{volume}{17}.
\newblock \bibinfo{publisher}{Princeton university press}.
\newblock


\bibitem[\protect\citeauthoryear{Minsky}{Minsky}{1988}]%
        {minsky1988society}
\bibfield{author}{\bibinfo{person}{Marvin Minsky}.}
  \bibinfo{year}{1988}\natexlab{}.
\newblock \bibinfo{booktitle}{\emph{Society of mind}}.
\newblock \bibinfo{publisher}{Simon and Schuster}.
\newblock


\bibitem[\protect\citeauthoryear{Moltz}{Moltz}{1965}]%
        {moltz1965contemporary}
\bibfield{author}{\bibinfo{person}{Howard Moltz}.}
  \bibinfo{year}{1965}\natexlab{}.
\newblock \showarticletitle{Contemporary instinct theory and the fixed action
  pattern.}
\newblock \bibinfo{journal}{\emph{Psychological Review}} \bibinfo{volume}{72},
  \bibinfo{number}{1} (\bibinfo{year}{1965}), \bibinfo{pages}{27}.
\newblock


\bibitem[\protect\citeauthoryear{Munar, Hovmand, Fleming, and Darmstadt}{Munar
  et~al\mbox{.}}{2015}]%
        {munar2015scaling}
\bibfield{author}{\bibinfo{person}{Wolfgang Munar}, \bibinfo{person}{Peter~S
  Hovmand}, \bibinfo{person}{Carrie Fleming}, {and} \bibinfo{person}{Gary~L
  Darmstadt}.} \bibinfo{year}{2015}\natexlab{}.
\newblock \showarticletitle{Scaling-up impact in perinatology through systems
  science: Bridging the collaboration and translational divides in
  cross-disciplinary research and public policy}. In
  \bibinfo{booktitle}{\emph{Seminars in perinatology}},
  Vol.~\bibinfo{volume}{39}. Elsevier, \bibinfo{pages}{416--423}.
\newblock


\bibitem[\protect\citeauthoryear{Noble and Rittel}{Noble and Rittel}{1988}]%
        {noble1988issue}
\bibfield{author}{\bibinfo{person}{Douglas Noble} {and}
  \bibinfo{person}{Horst~WJ Rittel}.} \bibinfo{year}{1988}\natexlab{}.
\newblock \showarticletitle{Issue-based information systems for design}.
\newblock  (\bibinfo{year}{1988}).
\newblock


\bibitem[\protect\citeauthoryear{Osoba and Kosko}{Osoba and Kosko}{2019}]%
        {osoba2019beyond}
\bibfield{author}{\bibinfo{person}{Osonde Osoba} {and} \bibinfo{person}{Bart
  Kosko}.} \bibinfo{year}{2019}\natexlab{}.
\newblock \showarticletitle{Beyond DAGs: Modeling Causal Feedback with Fuzzy
  Cognitive Maps}.
\newblock \bibinfo{journal}{\emph{arXiv preprint arXiv:1906.11247}}
  (\bibinfo{year}{2019}).
\newblock


\bibitem[\protect\citeauthoryear{Page}{Page}{2018}]%
        {page2018model}
\bibfield{author}{\bibinfo{person}{Scott~E Page}.}
  \bibinfo{year}{2018}\natexlab{}.
\newblock \bibinfo{booktitle}{\emph{The Model Thinker: What You Need to Know to
  Make Data Work for You}}.
\newblock \bibinfo{publisher}{Hachette UK}.
\newblock


\bibitem[\protect\citeauthoryear{Pearl}{Pearl}{2009}]%
        {pearl2009causality}
\bibfield{author}{\bibinfo{person}{Judea Pearl}.}
  \bibinfo{year}{2009}\natexlab{}.
\newblock \bibinfo{booktitle}{\emph{Causality}}.
\newblock \bibinfo{publisher}{Cambridge university press}.
\newblock


\bibitem[\protect\citeauthoryear{Pearl and Mackenzie}{Pearl and
  Mackenzie}{2018}]%
        {pearl2018book}
\bibfield{author}{\bibinfo{person}{Judea Pearl} {and} \bibinfo{person}{Dana
  Mackenzie}.} \bibinfo{year}{2018}\natexlab{}.
\newblock \bibinfo{booktitle}{\emph{The book of why: the new science of cause
  and effect}}.
\newblock \bibinfo{publisher}{Basic Books}.
\newblock


\bibitem[\protect\citeauthoryear{Peck}{Peck}{1998}]%
        {peck1998group}
\bibfield{author}{\bibinfo{person}{S Peck}.} \bibinfo{year}{1998}\natexlab{}.
\newblock \showarticletitle{Group model building: facilitating team learning
  using system dynamics}.
\newblock \bibinfo{journal}{\emph{Journal of the Operational Research Society}}
  \bibinfo{volume}{49}, \bibinfo{number}{7} (\bibinfo{year}{1998}),
  \bibinfo{pages}{766--767}.
\newblock


\bibitem[\protect\citeauthoryear{Pinch and Bijker}{Pinch and Bijker}{1984}]%
        {pinch1984social}
\bibfield{author}{\bibinfo{person}{Trevor~J Pinch} {and}
  \bibinfo{person}{Wiebe~E Bijker}.} \bibinfo{year}{1984}\natexlab{}.
\newblock \showarticletitle{The social construction of facts and artefacts: Or
  how the sociology of science and the sociology of technology might benefit
  each other}.
\newblock \bibinfo{journal}{\emph{Social studies of science}}
  \bibinfo{volume}{14}, \bibinfo{number}{3} (\bibinfo{year}{1984}),
  \bibinfo{pages}{399--441}.
\newblock


\bibitem[\protect\citeauthoryear{Plsek and Greenhalgh}{Plsek and
  Greenhalgh}{2001}]%
        {plsek2001challenge}
\bibfield{author}{\bibinfo{person}{Paul~E Plsek} {and} \bibinfo{person}{Trisha
  Greenhalgh}.} \bibinfo{year}{2001}\natexlab{}.
\newblock \showarticletitle{The challenge of complexity in health care}.
\newblock \bibinfo{journal}{\emph{Bmj}} \bibinfo{volume}{323},
  \bibinfo{number}{7313} (\bibinfo{year}{2001}), \bibinfo{pages}{625--628}.
\newblock


\bibitem[\protect\citeauthoryear{Rabiee}{Rabiee}{2004}]%
        {rabiee2004focus}
\bibfield{author}{\bibinfo{person}{Fatemeh Rabiee}.}
  \bibinfo{year}{2004}\natexlab{}.
\newblock \showarticletitle{Focus-group interview and data analysis}.
\newblock \bibinfo{journal}{\emph{Proceedings of the nutrition society}}
  \bibinfo{volume}{63}, \bibinfo{number}{4} (\bibinfo{year}{2004}),
  \bibinfo{pages}{655--660}.
\newblock


\bibitem[\protect\citeauthoryear{Rahwan, Cebrian, Obradovich, Bongard,
  Bonnefon, Breazeal, Crandall, Christakis, Couzin, Jackson, Jennings, Kamar,
  Kloumann, Larochelle, Lazer, {McElreath}, Mislove, Parkes, Pentland, Roberts,
  Shariff, Tenenbaum, and Wellman}{Rahwan et~al\mbox{.}}{[n.d.]}]%
        {rahwan_machine_2019}
\bibfield{author}{\bibinfo{person}{Iyad Rahwan}, \bibinfo{person}{Manuel
  Cebrian}, \bibinfo{person}{Nick Obradovich}, \bibinfo{person}{Josh Bongard},
  \bibinfo{person}{Jean-François Bonnefon}, \bibinfo{person}{Cynthia
  Breazeal}, \bibinfo{person}{Jacob~W. Crandall}, \bibinfo{person}{Nicholas~A.
  Christakis}, \bibinfo{person}{Iain~D. Couzin}, \bibinfo{person}{Matthew~O.
  Jackson}, \bibinfo{person}{Nicholas~R. Jennings}, \bibinfo{person}{Ece
  Kamar}, \bibinfo{person}{Isabel~M. Kloumann}, \bibinfo{person}{Hugo
  Larochelle}, \bibinfo{person}{David Lazer}, \bibinfo{person}{Richard
  {McElreath}}, \bibinfo{person}{Alan Mislove}, \bibinfo{person}{David~C.
  Parkes}, \bibinfo{person}{Alex~`Sandy' Pentland},
  \bibinfo{person}{Margaret~E. Roberts}, \bibinfo{person}{Azim Shariff},
  \bibinfo{person}{Joshua~B. Tenenbaum}, {and} \bibinfo{person}{Michael
  Wellman}.} \bibinfo{year}{[n.d.]}\natexlab{}.
\newblock \showarticletitle{Machine behaviour}.
\newblock  \bibinfo{volume}{568}, \bibinfo{number}{7753}
  (\bibinfo{year}{[n.\,d.]}), \bibinfo{pages}{477--486}.
\newblock


\bibitem[\protect\citeauthoryear{Raso, Hilligoss, Krishnamurthy, Bavitz, and
  Kim}{Raso et~al\mbox{.}}{2018}]%
        {raso2018artificial}
\bibfield{author}{\bibinfo{person}{Filippo~A Raso}, \bibinfo{person}{Hannah
  Hilligoss}, \bibinfo{person}{Vivek Krishnamurthy},
  \bibinfo{person}{Christopher Bavitz}, {and} \bibinfo{person}{Levin Kim}.}
  \bibinfo{year}{2018}\natexlab{}.
\newblock \showarticletitle{Artificial Intelligence \& Human Rights:
  Opportunities \& Risks}.
\newblock \bibinfo{journal}{\emph{Berkman Klein Center Research Publication}}
  \bibinfo{number}{2018-6} (\bibinfo{year}{2018}).
\newblock


\bibitem[\protect\citeauthoryear{Repenning, Kieffer, and Astor}{Repenning
  et~al\mbox{.}}{2017}]%
        {repenning2017most}
\bibfield{author}{\bibinfo{person}{Nelson~Peter Repenning},
  \bibinfo{person}{Don Kieffer}, {and} \bibinfo{person}{Todd Astor}.}
  \bibinfo{year}{2017}\natexlab{}.
\newblock \bibinfo{booktitle}{\emph{The most underrated skill in management}}.
\newblock \bibinfo{publisher}{MIT Sloan Management Review}.
\newblock


\bibitem[\protect\citeauthoryear{Richardson}{Richardson}{1991}]%
        {richardson1991feedback}
\bibfield{author}{\bibinfo{person}{George~P Richardson}.}
  \bibinfo{year}{1991}\natexlab{}.
\newblock \bibinfo{booktitle}{\emph{Feedback thought in social science and
  systems theory}}.
\newblock \bibinfo{publisher}{University of Pennsylvania}.
\newblock


\bibitem[\protect\citeauthoryear{Richardson}{Richardson}{2011}]%
        {richardson2011reflections}
\bibfield{author}{\bibinfo{person}{George~P Richardson}.}
  \bibinfo{year}{2011}\natexlab{}.
\newblock \showarticletitle{Reflections on the foundations of system dynamics}.
\newblock \bibinfo{journal}{\emph{System Dynamics Review}}
  \bibinfo{volume}{27}, \bibinfo{number}{3} (\bibinfo{year}{2011}),
  \bibinfo{pages}{219--243}.
\newblock


\bibitem[\protect\citeauthoryear{Richardson, Andersen, Maxwell, and
  Stewart}{Richardson et~al\mbox{.}}{1994}]%
        {richardson1994foundations}
\bibfield{author}{\bibinfo{person}{George~P Richardson},
  \bibinfo{person}{David~F Andersen}, \bibinfo{person}{Terrence~A Maxwell},
  {and} \bibinfo{person}{Thomas~R Stewart}.} \bibinfo{year}{1994}\natexlab{}.
\newblock \showarticletitle{Foundations of mental model research}. In
  \bibinfo{booktitle}{\emph{Proceedings of the 1994 International System
  Dynamics Conference}}. EF Wolstenholme, \bibinfo{pages}{181--192}.
\newblock


\bibitem[\protect\citeauthoryear{Richardson, Schultz, and Crawford}{Richardson
  et~al\mbox{.}}{2019}]%
        {richardson2019dirty}
\bibfield{author}{\bibinfo{person}{Rashida Richardson}, \bibinfo{person}{Jason
  Schultz}, {and} \bibinfo{person}{Kate Crawford}.}
  \bibinfo{year}{2019}\natexlab{}.
\newblock \showarticletitle{Dirty Data, Bad Predictions: How Civil Rights
  Violations Impact Police Data, Predictive Policing Systems, and Justice}.
\newblock \bibinfo{journal}{\emph{New York University Law Review Online,
  Forthcoming}} (\bibinfo{year}{2019}).
\newblock


\bibitem[\protect\citeauthoryear{Richmond}{Richmond}{1993}]%
        {richmond1993systems}
\bibfield{author}{\bibinfo{person}{Barry Richmond}.}
  \bibinfo{year}{1993}\natexlab{}.
\newblock \showarticletitle{Systems thinking: critical thinking skills for the
  1990s and beyond}.
\newblock \bibinfo{journal}{\emph{System dynamics review}} \bibinfo{volume}{9},
  \bibinfo{number}{2} (\bibinfo{year}{1993}), \bibinfo{pages}{113--133}.
\newblock


\bibitem[\protect\citeauthoryear{Rouse}{Rouse}{2008}]%
        {rouse2008health}
\bibfield{author}{\bibinfo{person}{William~B Rouse}.}
  \bibinfo{year}{2008}\natexlab{}.
\newblock \showarticletitle{Health care as a complex adaptive system:
  implications for design and management}.
\newblock \bibinfo{journal}{\emph{Bridge-Washington-National Academy of
  Engineering-}} \bibinfo{volume}{38}, \bibinfo{number}{1}
  (\bibinfo{year}{2008}), \bibinfo{pages}{17}.
\newblock


\bibitem[\protect\citeauthoryear{Saeed}{Saeed}{1992}]%
        {saeed1992slicing}
\bibfield{author}{\bibinfo{person}{Khalid Saeed}.}
  \bibinfo{year}{1992}\natexlab{}.
\newblock \showarticletitle{Slicing a complex problem for system dynamics
  modeling}.
\newblock \bibinfo{journal}{\emph{System Dynamics Review}} \bibinfo{volume}{8},
  \bibinfo{number}{3} (\bibinfo{year}{1992}), \bibinfo{pages}{251--261}.
\newblock


\bibitem[\protect\citeauthoryear{Saeed}{Saeed}{1998}]%
        {saeed1998defining}
\bibfield{author}{\bibinfo{person}{Khalid Saeed}.}
  \bibinfo{year}{1998}\natexlab{}.
\newblock \bibinfo{booktitle}{\emph{Defining a problem or constructing a
  reference mode}}.
\newblock \bibinfo{publisher}{Department of Social Science and Policy Studies,
  Worcester Polytechnic Institute}.
\newblock


\bibitem[\protect\citeauthoryear{Saysel, Barlas, and Yenig{\"u}n}{Saysel
  et~al\mbox{.}}{2002}]%
        {saysel2002environmental}
\bibfield{author}{\bibinfo{person}{Ali~Kerem Saysel}, \bibinfo{person}{Yaman
  Barlas}, {and} \bibinfo{person}{Orhan Yenig{\"u}n}.}
  \bibinfo{year}{2002}\natexlab{}.
\newblock \showarticletitle{Environmental sustainability in an agricultural
  development project: a system dynamics approach}.
\newblock \bibinfo{journal}{\emph{Journal of environmental management}}
  \bibinfo{volume}{64}, \bibinfo{number}{3} (\bibinfo{year}{2002}),
  \bibinfo{pages}{247--260}.
\newblock


\bibitem[\protect\citeauthoryear{Schneider and Somers}{Schneider and
  Somers}{2006}]%
        {schneider2006organizations}
\bibfield{author}{\bibinfo{person}{Marguerite Schneider} {and}
  \bibinfo{person}{Mark Somers}.} \bibinfo{year}{2006}\natexlab{}.
\newblock \showarticletitle{Organizations as complex adaptive systems:
  Implications of complexity theory for leadership research}.
\newblock \bibinfo{journal}{\emph{The Leadership Quarterly}}
  \bibinfo{volume}{17}, \bibinfo{number}{4} (\bibinfo{year}{2006}),
  \bibinfo{pages}{351--365}.
\newblock


\bibitem[\protect\citeauthoryear{Schwab}{Schwab}{2017}]%
        {schwab2017fourth}
\bibfield{author}{\bibinfo{person}{Klaus Schwab}.}
  \bibinfo{year}{2017}\natexlab{}.
\newblock \bibinfo{booktitle}{\emph{The fourth industrial revolution}}.
\newblock \bibinfo{publisher}{Crown Business}.
\newblock


\bibitem[\protect\citeauthoryear{Searle, Willis, et~al\mbox{.}}{Searle
  et~al\mbox{.}}{1995}]%
        {searle1995construction}
\bibfield{author}{\bibinfo{person}{John~R Searle}, \bibinfo{person}{S Willis},
  {et~al\mbox{.}}} \bibinfo{year}{1995}\natexlab{}.
\newblock \bibinfo{booktitle}{\emph{The construction of social reality}}.
\newblock \bibinfo{publisher}{Simon and Schuster}.
\newblock


\bibitem[\protect\citeauthoryear{Selbst, Friedler, Venkatasubramanian, Vertesi,
  et~al\mbox{.}}{Selbst et~al\mbox{.}}{2018}]%
        {selbst2018fairness}
\bibfield{author}{\bibinfo{person}{Andrew~D Selbst}, \bibinfo{person}{Sorelle
  Friedler}, \bibinfo{person}{Suresh Venkatasubramanian},
  \bibinfo{person}{Janet Vertesi}, {et~al\mbox{.}}}
  \bibinfo{year}{2018}\natexlab{}.
\newblock \showarticletitle{Fairness and Abstraction in Sociotechnical
  Systems}. In \bibinfo{booktitle}{\emph{ACM Conference on Fairness,
  Accountability, and Transparency (FAT*)}}.
\newblock


\bibitem[\protect\citeauthoryear{Shrier and Platt}{Shrier and Platt}{2008}]%
        {shrier2008reducing}
\bibfield{author}{\bibinfo{person}{Ian Shrier} {and} \bibinfo{person}{Robert~W
  Platt}.} \bibinfo{year}{2008}\natexlab{}.
\newblock \showarticletitle{Reducing bias through directed acyclic graphs}.
\newblock \bibinfo{journal}{\emph{BMC medical research methodology}}
  \bibinfo{volume}{8}, \bibinfo{number}{1} (\bibinfo{year}{2008}),
  \bibinfo{pages}{70}.
\newblock


\bibitem[\protect\citeauthoryear{Simon}{Simon}{1982}]%
        {simon1982models}
\bibfield{author}{\bibinfo{person}{Herbert~Alexander Simon}.}
  \bibinfo{year}{1982}\natexlab{}.
\newblock \bibinfo{booktitle}{\emph{Models of bounded rationality: Empirically
  grounded economic reason}}. Vol.~\bibinfo{volume}{3}.
\newblock \bibinfo{publisher}{Massachussetts Institute of Technology Press}.
\newblock


\bibitem[\protect\citeauthoryear{Stave and Kopainsky}{Stave and
  Kopainsky}{2015}]%
        {stave2015system}
\bibfield{author}{\bibinfo{person}{Krystyna~A Stave} {and}
  \bibinfo{person}{Birgit Kopainsky}.} \bibinfo{year}{2015}\natexlab{}.
\newblock \showarticletitle{A system dynamics approach for examining mechanisms
  and pathways of food supply vulnerability}.
\newblock \bibinfo{journal}{\emph{Journal of Environmental Studies and
  Sciences}} \bibinfo{volume}{5}, \bibinfo{number}{3} (\bibinfo{year}{2015}),
  \bibinfo{pages}{321--336}.
\newblock


\bibitem[\protect\citeauthoryear{Sterman}{Sterman}{2000}]%
        {sterman2000business}
\bibfield{author}{\bibinfo{person}{John~D. Sterman}.}
  \bibinfo{year}{2000}\natexlab{}.
\newblock \bibinfo{booktitle}{\emph{Business dynamics: Systems thinking and
  modeling for a complex world}}.
\newblock \bibinfo{publisher}{McGraw-Hill}.
\newblock


\bibitem[\protect\citeauthoryear{Sterman}{Sterman}{2001}]%
        {sterman2001system}
\bibfield{author}{\bibinfo{person}{John~D. Sterman}.}
  \bibinfo{year}{2001}\natexlab{}.
\newblock \showarticletitle{System dynamics modeling: tools for learning in a
  complex world}.
\newblock \bibinfo{journal}{\emph{California management review}}
  \bibinfo{volume}{43}, \bibinfo{number}{4} (\bibinfo{year}{2001}),
  \bibinfo{pages}{8--25}.
\newblock


\bibitem[\protect\citeauthoryear{Sterman}{Sterman}{2006}]%
        {sterman2006learning}
\bibfield{author}{\bibinfo{person}{John~D. Sterman}.}
  \bibinfo{year}{2006}\natexlab{}.
\newblock \showarticletitle{Learning from evidence in a complex world}.
\newblock \bibinfo{journal}{\emph{American journal of public health}}
  \bibinfo{volume}{96}, \bibinfo{number}{3} (\bibinfo{year}{2006}),
  \bibinfo{pages}{505--514}.
\newblock


\bibitem[\protect\citeauthoryear{Surana, Kumara*, Greaves, and Raghavan}{Surana
  et~al\mbox{.}}{2005}]%
        {surana2005supply}
\bibfield{author}{\bibinfo{person}{Amit Surana}, \bibinfo{person}{Soundar
  Kumara*}, \bibinfo{person}{Mark Greaves}, {and} \bibinfo{person}{Usha~Nandini
  Raghavan}.} \bibinfo{year}{2005}\natexlab{}.
\newblock \showarticletitle{Supply-chain networks: a complex adaptive systems
  perspective}.
\newblock \bibinfo{journal}{\emph{International Journal of Production
  Research}} \bibinfo{volume}{43}, \bibinfo{number}{20} (\bibinfo{year}{2005}),
  \bibinfo{pages}{4235--4265}.
\newblock


\bibitem[\protect\citeauthoryear{Tenenbaum and Griffiths}{Tenenbaum and
  Griffiths}{2003}]%
        {tenenbaum2003theory}
\bibfield{author}{\bibinfo{person}{Joshua~B Tenenbaum} {and}
  \bibinfo{person}{Thomas~L Griffiths}.} \bibinfo{year}{2003}\natexlab{}.
\newblock \showarticletitle{Theory-based causal inference}. In
  \bibinfo{booktitle}{\emph{Advances in neural information processing
  systems}}. \bibinfo{pages}{43--50}.
\newblock


\bibitem[\protect\citeauthoryear{Thompsonab and Tebbensc}{Thompsonab and
  Tebbensc}{2008}]%
        {thompsonab2008using}
\bibfield{author}{\bibinfo{person}{Kimberly~M Thompsonab} {and}
  \bibinfo{person}{Radboud J~Duintjer Tebbensc}.}
  \bibinfo{year}{2008}\natexlab{}.
\newblock \showarticletitle{Using system dynamics to develop policies that
  matter: global management of poliomyelitis and beyond}.
\newblock \bibinfo{journal}{\emph{System Dynamics Review}}
  \bibinfo{volume}{24}, \bibinfo{number}{4} (\bibinfo{year}{2008}),
  \bibinfo{pages}{433--449}.
\newblock


\bibitem[\protect\citeauthoryear{Trani, Bakhshi, Mozaffari, Sohail, Rawab,
  Kaplan, Ballard, and Hovmand}{Trani et~al\mbox{.}}{2019}]%
        {trani2019strengthening}
\bibfield{author}{\bibinfo{person}{Jean-Fran{\c{c}}ois Trani},
  \bibinfo{person}{Parul Bakhshi}, \bibinfo{person}{Alan Mozaffari},
  \bibinfo{person}{Munib Sohail}, \bibinfo{person}{Hashim Rawab},
  \bibinfo{person}{Ian Kaplan}, \bibinfo{person}{Ellis Ballard}, {and}
  \bibinfo{person}{Peter Hovmand}.} \bibinfo{year}{2019}\natexlab{}.
\newblock \showarticletitle{Strengthening child inclusion in the classroom in
  rural schools of Pakistan and Afghanistan: What did we learn by testing the
  system dynamics protocol for community engagement?}
\newblock \bibinfo{journal}{\emph{Research in Comparative and International
  Education}} \bibinfo{volume}{14}, \bibinfo{number}{1} (\bibinfo{year}{2019}),
  \bibinfo{pages}{158--181}.
\newblock


\bibitem[\protect\citeauthoryear{Trani, Ballard, Bakhshi, and Hovmand}{Trani
  et~al\mbox{.}}{2016}]%
        {trani2016community}
\bibfield{author}{\bibinfo{person}{Jean-Francois Trani}, \bibinfo{person}{Ellis
  Ballard}, \bibinfo{person}{Parul Bakhshi}, {and} \bibinfo{person}{Peter
  Hovmand}.} \bibinfo{year}{2016}\natexlab{}.
\newblock \showarticletitle{Community based system dynamic as an approach for
  understanding and acting on messy problems: a case study for global mental
  health intervention in Afghanistan}.
\newblock \bibinfo{journal}{\emph{Conflict and health}} \bibinfo{volume}{10},
  \bibinfo{number}{1} (\bibinfo{year}{2016}), \bibinfo{pages}{25}.
\newblock


\bibitem[\protect\citeauthoryear{Von~Bertalanffy}{Von~Bertalanffy}{1950}]%
        {von1950theory}
\bibfield{author}{\bibinfo{person}{Ludwig Von~Bertalanffy}.}
  \bibinfo{year}{1950}\natexlab{}.
\newblock \showarticletitle{The theory of open systems in physics and biology}.
\newblock \bibinfo{journal}{\emph{Science}} \bibinfo{volume}{111},
  \bibinfo{number}{2872} (\bibinfo{year}{1950}), \bibinfo{pages}{23--29}.
\newblock


\bibitem[\protect\citeauthoryear{Whittaker, Crawford, Dobbe, Fried, Kaziunas,
  Mathur, West, Richardson, Schultz, and Schwartz}{Whittaker
  et~al\mbox{.}}{2018}]%
        {whittaker2018ai}
\bibfield{author}{\bibinfo{person}{Meredith Whittaker}, \bibinfo{person}{Kate
  Crawford}, \bibinfo{person}{Roel Dobbe}, \bibinfo{person}{Genevieve Fried},
  \bibinfo{person}{Elizabeth Kaziunas}, \bibinfo{person}{Varoon Mathur},
  \bibinfo{person}{Sarah~Mysers West}, \bibinfo{person}{Rashida Richardson},
  \bibinfo{person}{Jason Schultz}, {and} \bibinfo{person}{Oscar Schwartz}.}
  \bibinfo{year}{2018}\natexlab{}.
\newblock \bibinfo{booktitle}{\emph{AI now report 2018}}.
\newblock \bibinfo{publisher}{AI Now Institute at New York University}.
\newblock


\bibitem[\protect\citeauthoryear{Yeung}{Yeung}{2011}]%
        {yeung2011investigation}
\bibfield{author}{\bibinfo{person}{Sai~Wing Yeung}.}
  \bibinfo{year}{2011}\natexlab{}.
\newblock \emph{\bibinfo{title}{An investigation of human inductive biases in
  causality and probability judgments}}.
\newblock \bibinfo{thesistype}{Ph.D. Dissertation}. \bibinfo{school}{UC
  Berkeley}.
\newblock


\bibitem[\protect\citeauthoryear{Young, Magassa, and Friedman}{Young
  et~al\mbox{.}}{2019}]%
        {young2019toward}
\bibfield{author}{\bibinfo{person}{Meg Young}, \bibinfo{person}{Lassana
  Magassa}, {and} \bibinfo{person}{Batya Friedman}.}
  \bibinfo{year}{2019}\natexlab{}.
\newblock \showarticletitle{Toward inclusive tech policy design: a method for
  underrepresented voices to strengthen tech policy documents}.
\newblock \bibinfo{journal}{\emph{Ethics and Information Technology}}
  \bibinfo{volume}{21}, \bibinfo{number}{2} (\bibinfo{year}{2019}),
  \bibinfo{pages}{89--103}.
\newblock


\end{thebibliography}

\newpage
\appendix

\onecolumn

\section*{Appendix}
\subsection*{Model Variables and Underlying Equations}





We have placed all models used in this paper at this anonymous online drive (\url{https://figshare.com/s/1d341bf4e24815d7db99}) which can be inspected (i.e., simulating using different settings) using the free Stella Player software (\url{https://www.iseesystems.com/softwares/player/iseeplayer.aspx}).
The descriptions of different variables, and their associated types and units are shown in Table~\ref{tab_vars}.
Then, for $j\in \{A, B\}$, where $A$ and $B$ are two population groups, we define the various terms, their initial values and their relationships as follows:

\begin{align*}
    A^j(t) &=A^j_0 + \int\limits_0^t (n(t) - d(t))\;dt \\
    O^j(t) &=O^j_0 + \int\limits_0^t (r(t) - p(t) -f(t))\;dt \\
    r^j(t) &=\alpha\times \tau \times g^j(t)\\
    p^j(t) &=O^j(t)\times x^j\\
    f^j(t) &=O^j(t)\times (1-x^j)\\
    n^j(t) &=\frac{O^j\times \iota \times \sigma}{\tau}\times \frac{\sigma - S(t)}{\sigma}\\
    d^j(t) &=\frac{A^j(t)\times f(t) \times \delta}{\tau}\\
    \upsilon^j(t) &=
    \begin{cases}
    10 \text{ if } (t) < 10 &\text{Under adjusting loan term intervention}\\
    10\times x \text{ otherwise}
    \end{cases}\\
       \lambda^j(t) &=
    \begin{cases}
   400 \text{ if } (t) < 10 &\text{Under lowering credit score threshold intervention}\\
    200 \text{ otherwise}
    \end{cases}\\
    x^j(t) &= \left(\frac{1}{1+\exp(-(47.89-0.083\times A^j(t)))}-1\right)\times 4+5 \\
    g^j(t) &= \frac{1}{1+\exp(-(3.57+3.43*A^j(t)/\lambda))} \\
    \lambda &= 400\\
    \alpha &= .5\\
    \tau &= 10000\\
    \upsilon &= 10\\
    \pi^j &= 
    \begin{cases}
    0.8 \text{ if } j = A\\
    0.6 \text{ if } j = B\\
    \end{cases}\\
    \sigma &= 850\\
    \iota &= .04\\
    \delta &= .25\\
\end{align*}

\noindent\textbf{Simulation Settings:}

\begin{itemize}
    \item Simulation start time = 0
    \item Simulation end time = 20
    \item Time units = years
    \item DT = 1/12
\end{itemize}

\begin{table}[th]
\begin{tabular}{@{}llll@{}}
\toprule
\textbf{Variable}                      & \textbf{Type} & \textbf{Units}     & \textbf{Symbol} \\ \midrule
average credit score                   & stock         & points/people           & $S$             \\
borrowers                              & stock         & people             & $O$             \\
receiving loans                        & flow          & people/year        & $r$             \\
paying off                             & flow          & people/year        & $p$             \\
defaulting                             & flow          & people/year        & $f$             \\
increasing credit score                & flow          & points/year        & $n$             \\
decreasing credit score                & flow          & points/year        & $d$             \\
loan granting threshold                & parameter     & points             & $\lambda$       \\
application rate                       & parameter     & dimensionless/year & $\alpha$        \\
total population                       & parameter     & people             & $\tau$          \\
average loan term                      & parameter     & year               & $\upsilon$      \\
probability of payoff          & parameter     & dimensionless      & $\pi^A$, $\pi^B$           \\
max credit score                       & parameter     & points             & $\sigma$        \\ 
average increase per year of repayment & parameter     & dimensionless $\times$ people /year & $\iota$         \\
average decrease per default & parameter     & dimensionless $\times$ people/year & $\delta$        \\
fraction of loans granted               & auxiliary     & dimensionless      & $g$             \\
interest rate                          & auxiliary     & dimensionless      & $i$             \\
effect of credit score on loan term length                         & auxiliary     & dimensionless      & $x$             \\
\bottomrule
\end{tabular}
\caption{Description of system variables used for the model described in Figure~\ref{fig_sd_loans_moritz}.\label{tab_vars}}
\end{table}


\end{document}